\documentclass[preprint,12pt]{aastex}

\shorttitle{Implications of Early Afterglows of Swift GRBs}

\begin{document}

\title{Implications of the Early X-Ray Afterglow Light Curves of Swift
GRBs}

\author{Jonathan Granot\altaffilmark{1}, Arieh K\"onigl\altaffilmark{2},
  Tsvi Piran\altaffilmark{3}}

\altaffiltext{1}{KIPAC, Stanford University, P.O. Box 20450, MS
29, Stanford, CA 94309; granot@slac.stanford.edu}
\altaffiltext{2}{Department of Astronomy and Astrophysics and
Enrico Fermi Institute, University of Chicago, 5640 South Ellis
Avenue, Chicago, IL 60637; arieh@jets.uchicago.edu}
\altaffiltext{3}{Racah Institute of Physics, The Hebrew
University, Jerusalem, 91904, Israel; tsvi@phys.huji.ac.il}

\begin{abstract}

According to current models, gamma-ray bursts (GRBs) are produced when
the energy carried by a relativistic outflow is dissipated and
converted into radiation. The efficiency of this process,
$\epsilon_\gamma$, is one of the critical factors in any GRB
model. The X-ray afterglow light curves of {\it Swift}\/ GRBs show an
early stage of flattish decay. This has been interpreted as reflecting
energy injection. When combined with previous estimates, which have
concluded that the kinetic energy of the late ($\gtrsim 10\;$hr)
afterglow is comparable to the energy emitted in $\gamma$-rays, this
interpretation implies very high values of $\epsilon_\gamma$,
corresponding to $\gtrsim 90\%$ of the initial energy being converted
into $\gamma$-rays. Such a high efficiency is hard to reconcile with
most models, including in particular the popular internal-shocks
model. We re-analyze the derivation of the kinetic energy from the
afterglow X-ray flux and re-examine the resulting estimates of the
efficiency. We confirm that, if the flattish decay arises from energy
injection and the pre-{\it Swift}\/ broad-band estimates of the
kinetic energy are correct, then $\epsilon_\gamma\gtrsim 0.9$. We
discuss various issues related to this result, including an
alternative interpretation of the light curve in terms of a
two-component outflow model, which we apply to the X-ray observations
of GRB 050315.  We point out, however, that another interpretation of
the flattish decay --- a variable X-ray afterglow efficiency (e.g.,
due to a time dependence of afterglow shock microphysical parameters)
--- is possible. We also show that direct estimates of the kinetic
energy from the late X-ray afterglow flux are sensitive to the assumed
values of the shock microphysical parameters and suggest that
broad-band afterglow fits might have underestimated the kinetic energy
(e.g., by overestimating the fraction of electrons that are
accelerated to relativistic energies).  Either one of these
possibilities implies a lower $\gamma$-ray efficiency, and their joint
effect could conceivably reduce the estimate of the typical
$\epsilon_\gamma$ to a value in the range $\sim 0.1-0.5$.

\end{abstract}

\keywords{gamma-rays: bursts --- ISM: jets and outflows --- radiation
mechanisms: nonthermal --- X-rays: individual (GRB 050315)}

\section{Introduction}
\label{sec:introduction}

Recent observations by the {\it Swift}\/ X-ray telescope have provided new
information on the early behavior of the X-ray light curve of
long-duration ($\gtrsim 2\; {\rm s}$) gamma-ray burst (GRB)
sources. Specifically, it was found \citep{Nousek05} that the light
curves of these sources have a generic shape consisting of three distinct
power-law segments $\propto t^{-\alpha}$: an initial (at $t<t_{\rm
break,1}$, with $300\; {\rm s}\lesssim t_{\rm break,1} \lesssim 500\;
{\rm s}$) very steep decline with time $t$ (with a power-law index
$\alpha_1$ in the range $3\lesssim \alpha_1 \lesssim 5$; see also
\citealt{Tagliaferri05} and \citealt{Barthelmy05}); a subsequent (at
$t_{\rm break,1}<t<t_{\rm break,2}$, with $10^3\; {\rm s}\lesssim t_{\rm
break,2} \lesssim 10^4\; {\rm s}$) very shallow decay ($0.2\lesssim
\alpha_2 \lesssim 0.8$); and a final steepening (at $t>t_{\rm break,2}$)
to the canonical power-law behavior ($1\lesssim \alpha_3 \lesssim 1.5$)
that was known from pre-{\it Swift}\/ observations.

\citet{Nousek05} already recognized that these results have direct
consequences to the question of the $\gamma$-ray emission efficiency
in GRB sources. This question is important to our understanding of the
basic prompt-emission mechanism. In the currently accepted
interpretation \citep[e.g.,][]{P99,P04}, the $\gamma$-rays originate
in a relativistic jet that is launched from the vicinity of a newly
born neutron star or stellar-mass black hole. In the simplest picture,
a fraction $\epsilon_\gamma$ of the energy injected at the source is
given to the prompt radiation, with the remaining fraction
($1-\epsilon_\gamma$) ending up as kinetic energy of ambient gas that
is swept up by a forward shock and then mostly radiated as early
afterglow emission. One attractive mechanism for explaining the prompt
emission characteristics invokes internal shocks that are generated
when the ejecta have a nonuniform distribution of Lorentz factors,
which results in outflowing ``shells'' colliding with each other at
large distances from the source.

Pre-{\it Swift}\/ observations, based on measurements of the
$\gamma$-ray fluence and of the ``late'' ($\gtrsim 10\; {\rm hr}$)
afterglow emission, have implied (when interpreted in the context of
the basic jet model) comparable (and narrowly clustered) values for
the radiated $\gamma$-ray energy and the kinetic energy feeding the
afterglow emission, i.e., $\epsilon_\gamma \sim 0.5$
\citep[e.g.,][]{F01,PK01a,PK01b,PK02,BKF03,BFK03,Y03}. This result is
seemingly problematic for the internal-shocks model, for which an
order-of-magnitude lower value for $\epsilon_\gamma$ is a more natural
expectation \citep{KPS97,DM98,K99,GSW01}. This apparent difficulty
could in principle be overcome if the ejected shells have a highly
nonuniform distribution of Lorentz factors \citep[e.g.,][]{B00,KS01}.
However, to fit the data the shells must also satisfy a number of
other restrictive conditions, which reduces the attractiveness of this
interpretation (see \S~\ref{sec:conclusion}). An
alternative proposal was made by \citet{PKG05}, who argued that if the
jet consists of an ultra-relativistic narrow component (from which the
prompt emission originates) and a moderately relativistic wide
component, with the latter having a higher kinetic energy and the
former a higher kinetic energy per unit solid angle, then the wide
component would dominate the late afterglow emission and the
$\gamma$-ray radiative efficiency of the narrow component could be
significantly lower than the value of $\epsilon_\gamma$ inferred under
the assumption of a single-component jet. As explained in
\citet{PKG05}, this proposal was motivated by observational
indications of the presence of two components in the late afterglow
light curves of several GRB sources and by the predictions of certain
GRB source models.

The earliest (steepest) segment of the afterglow light curve is most
naturally explained as radiation at large angles to our line of sight
corresponding either to the prompt emission \citep{KP00a} or to
emission from the reverse shock that is driven into the ejecta
\citep{Kobayashi05}. This implies that the early afterglow emission is
much weaker than what would be expected on the basis of an
extrapolation from the late-afterglow data.  This behavior was
interpreted by \citet{Nousek05} as an indication of an even higher
$\gamma$-ray emission efficiency, typically $\epsilon_\gamma \sim
0.9$. Such a high efficiency could render the internal-shocks model
untenable.

Our primary goal is to evaluate the prompt-emission efficiency as
accurately and systematically as possible on the basis of current
data. For this purpose we re-derive in \S~\ref{sec:expressions}
expressions that explicitly relate $\epsilon_\gamma$ to observable
quantities. In particular, we express $\epsilon_\gamma$ in terms of
the product $\kappa f$ of two parameters, one ($\kappa$)
encapsulating information that could be obtained by pre-{\it Swift}\/
observations, and the other ($f$) representing early-time data
obtained in {\it Swift} measurements. In \S~\ref{sec:ek_est} we
re-examine the estimates of the kinetic energy during the afterglow
phase as inferred from the X-ray flux and present a new general
formulation, correcting errors that have propagated in the literature
and have generally led to an underestimate of the kinetic energy. We
then analyze both pre-{\it Swift}\/ (\S~\ref{sec:pre_swift}) and {\it
Swift} (\S~\ref{sec:swift}) data in a uniform manner in the context of
this formalism.  We argue that the kinetic energy estimates remain
subject to considerable uncertainties.  In particular, while the
simple analysis of a large number of bursts (using ``typical'' values
for the microphysical parameters) yields rather high values for the
kinetic energy and hence a low inferred $\gamma$-ray efficiency, a
multiwavelength analysis (which determines the microphysical
parameters from the fit to the data) of a small subset suggests that
the kinetic energy is lower and hence the inferred value of
$\epsilon_\gamma$ is higher. Some caveats to this analysis are
considered in \S~\ref{sec:assess}.

The suggestion that the {\it Swift}\/ observations imply a higher
value of $\epsilon_\gamma$ in comparison with the pre-{\it Swift}
results is based on the interpretation of the flattish segment of
the X-ray light curve as reflecting an increase in the kinetic
energy of the forward shock during the early stages of the
afterglow. In this picture, the kinetic energy just after the
prompt-emission phase was significantly lower than the kinetic
energy estimated from the later stages of the afterglow (the
pre-{\it Swift}\/ results). One conceivable way of avoiding the need
to increase the estimate of $\epsilon_\gamma$ in light of the new {\it Swift}\/
observations is through a time evolution (specifically, an
increase with $t$) of the X-ray afterglow emission efficiency
$\epsilon_{\rm X}$. Such a behavior could in principle also
account for the flattish segment of the light curve and eliminate
the need to invoke an increase in the shock kinetic energy. If, in
addition, the value of the afterglow kinetic energy at late times
were underestimated by the broad-band fits to pre-{\it Swift}\/
GRB afterglows, which could be the case if only a fraction
$\xi_e<1$ of the electrons were accelerated to relativistic
energies in the afterglow shock \citep[see][]{EW05}, then the
typical afterglow efficiency would be further reduced (to a value
as low as $\epsilon_\gamma\sim 0.1$ if $\xi_e \sim 0.1$), which
might reconcile the new {\it Swift}\/ data with the comparatively
low efficiencies expected in the internal-shocks model. We discuss
these issues in \S~\ref{sec:epsilon_X}.

A two-component jet model with the characteristics required for
reducing the inferred $\gamma$-ray emission efficiency is evidently
disfavored by the {\it Swift} data, but a model of this type with
different parameters could provide an alternative interpretation of
the flattish shape of the light curve between $t_{\rm break,1}$ and
$t_{\rm break,2}$. We elaborate on these matters in
\S~\ref{sec:2comp}, where we also present a tentative fit to the X-ray
light curve of the {\it Swift}\/ source GRB~050315 in the context of
this scenario.

Our conclusions on the physical implications of the early X-ray light-curve
observations of GRB sources are presented in
\S~\ref{sec:conclusion}.

\section{Estimating the Gamma-Ray Efficiency}
\label{sec:efficiency}

The observed isotropic-equivalent luminosity can generally be
expressed as $L_{\rm iso}=\epsilon E_{\rm iso}/T$, where $E_{\rm
iso}$ is the isotropic-equivalent energy of the relevant
component, $T$ is the duration of the relevant emission, and
$\epsilon$ is the overall efficiency. This efficiency is a
product of several factors: $\epsilon=\epsilon_{\rm
dis}\epsilon_e\epsilon_{\rm rad}\epsilon_{\rm obs}$, where a
fraction $\epsilon_{\rm dis}$ of the total energy is dissipated
into internal energy, a fraction $\epsilon_e$ of the internal
energy goes into electrons (or positrons) and can in principle
be radiated away, a fraction $\epsilon_{\rm rad}$ of the
electron energy is actually radiated, and a fraction
$\epsilon_{\rm obs}$ of the radiated energy falls within the
observed range of photon energies. A fraction
$\epsilon_\gamma=\epsilon_{\rm dis,GRB}\epsilon_{e,{\rm
GRB}}\epsilon_{\rm rad,GRB}$ of the total original (isotropic
equivalent) energy $E_{\rm iso,0}$ is radiated away during the
prompt emission, $E_{\rm\gamma,iso}=\epsilon_\gamma E_{\rm
iso,0}$, while the remaining (kinetic) energy in the
$\gamma$-ray emitting component of the outflow, $E_{\rm
k,iso,0}=(1-\epsilon_\gamma)E_{\rm iso,0}$, is transferred to
the afterglow shock at $t_{\rm dec}$. In the prompt GRB emission
$\epsilon_{\rm GRB}=\epsilon_\gamma\epsilon_{\rm obs,GRB}$. In
the afterglow $\epsilon_{\rm dis,X}\approx 1$, and therefore
$\epsilon_{\rm X}\approx\epsilon_{e,{\rm X}}\epsilon_{\rm
rad,X}\epsilon_{\rm obs,X}$, where we concentrate on the X-ray
afterglow.

Using the above expression for $\epsilon_\gamma$, it becomes clearer
why it is difficult for it to assume very high ($\gtrsim
0.9$) values. Whereas $\epsilon_{\rm rad,GRB}\approx 1$ is possible, and even
expected if the electrons cool significantly over a dynamical time
(as is typically expected in the internal-shocks model),
$\epsilon_\gamma\gtrsim 0.9$ requires in addition that both
$\epsilon_{\rm dis,GRB}>0.9$ and $\epsilon_{e,{\rm GRB}}>0.9$. It is
difficult to achieve $\epsilon_{\rm dis,GRB}>0.9$ (i.e., dissipate
more than 90\% of the total energy) in most models for the prompt
emission, and in particular in the internal-shocks model.
Furthermore, it is not trivial to produce $\epsilon_{e,{\rm GRB}}>0.9$
(i.e. more than 90\% of the postshock energy going to electrons) in
the internal-shocks model, where the electrons are believed to be
accelerated in a shock propagating into a magnetized proton-electron
plasma. In particular, this would require that less than
10\% of the postshock energy goes into the protons and the magnetic
field, which is difficult since the protons carry most of the energy
(in kinetic form) and the magnetic field likely carries a comparable
energy in the upstream fluid ahead of the shock. It might conceivably be
possible if less than 10\% of the energy is in the magnetic field and
the protons can somehow transfer their energy to the electrons, which
radiate it away.

\subsection{Relationship to Observed Quantities}
\label{sec:expressions}

The X-ray afterglow isotropic-equivalent luminosity, $L_{\rm X,iso}$,
can be expressed in terms of the X-ray afterglow flux, $F_{\rm X}$,
\begin{equation}\label{L_X}
L_{\rm X,iso}(t) = 4\pi d_L^2(1+z)^{\beta-\alpha-1} F_{\rm X}(t)\ ,
\end{equation}
if $F_\nu\propto\nu^{-\beta}t^{-\alpha}$ in the relevant ranges in
frequency and time.\footnote{For the more general case, see eq. [1] of
\citet{Nousek05} and the discussion thereafter. Here $L_{\rm
X,iso}(t)=\int_{\nu_1}^{\nu_2}d\nu\,L_{\nu,{\rm iso}}(t)$, where both
$\nu$ and $t$ are measured in the cosmological frame of the GRB, whereas
$F_{\rm X}(t)=\int_{\nu_1}^{\nu_2}d\nu\,F_\nu(t)$, where both $\nu$ and $t$
are measured in the observer frame.}  The efficiency of the X-ray
afterglow emission is defined as
\begin{equation}\label{epsilon_X}
\epsilon_{\rm X}(t)\equiv\frac{tL_{\rm X,iso}(t)}{E_{\rm k,iso}(t)} ,
\end{equation}
where $E_{\rm k,iso}$ is the isotropic-equivalent kinetic energy
in the afterglow shock.

Using the relation $E_{\rm iso,0} =
E_{\rm\gamma,iso}/\epsilon_\gamma = E_{\rm
k,iso,0}/(1-\epsilon_\gamma)$, we obtain
\begin{equation}\label{obs_ratio}
\frac{\epsilon_\gamma\epsilon_{\rm
obs,GRB}}{(1-\epsilon_\gamma)} = \frac{E_{\rm\gamma,iso}^{\rm
obs}}{E_{\rm k,iso,0}}=\kappa f \quad ,\quad
\kappa\equiv\frac{E_{\rm\gamma,iso}^{\rm obs}}{E_{\rm
k,iso}(t_*)} \quad ,\quad f\equiv\frac{E_{\rm
k,iso}(t_*)}{E_{\rm k,iso,0}}\ , \end{equation} where $t_*$ can
be chosen as a time at which it is convenient to estimate
$E_{\rm k,iso}$, and we shall use $t_* = 10\;$hr (since it is
widely used in the literature and is typically $>t_{\rm
break,2}$, the end of the flattish segment of the X-ray light
curve).

An important question is how to estimate $f$ and $\kappa$ from
observations.  The most straightforward way of estimating $f = E_{\rm
k,iso}(10\;{\rm hr})/E_{\rm k,iso,0}$, which has been used by
\citet{Nousek05}, is by evaluating the value of the X-ray flux
decrement at $t_{\rm dec}$ relative to the extrapolation to $t_{\rm
dec}$ of the late-time ($t>t_{\rm break,2}$) flux, and translating
this flux ratio into an energy ratio according to the standard
afterglow theory.  When estimating $f$ in this way we assume that
\begin{equation}\label{epsilon_X_standard}
\epsilon_{\rm X}\sim(1+Y)^{-1}\epsilon_e(\nu_m/\nu_{\rm X})^{(p-2)/2}\ ,
\end{equation}
where $\nu_m$ and $\nu_c$ are the characteristic synchrotron frequency
and cooling break frequency, respectively \citep{SPN98,GS02} and $Y$
is the Compton y-parameter. This result can also be obtained under the
assumptions of standard afterglow theory, as follows.  The overall
X-ray afterglow efficiency can be written as $\epsilon_{\rm X} \approx
\epsilon_e\epsilon_{\rm rad}\epsilon_{\rm obs}$, where $\epsilon_{\rm
rad} \approx \min[1,(\nu_m / \nu_c)^{(p-2)/2}]$ and $\epsilon_{\rm
obs} \approx (1+Y)^{-1}\max[(\nu_m/\nu_{\rm
X})^{(p-2)/2},(\nu_c/\nu_{\rm X})^{(p-2)/2}]$; the factor of
$(1+Y)^{-1}$ is the fraction of the radiated
energy in the synchrotron component, and it is present because
the synchrotron self-Compton (SSC)
component typically does not contribute significantly to $F_{\rm
X}(10\;{\rm hr})$ but may still dominate the total radiated
luminosity. The factor $(1+Y)^{-1}$ is generally assumed to be small,
consistent with the usual inference that the magnetic-to-thermal
energy ratio in the emission region, $\epsilon_B$, is smaller than
$\epsilon_e$ (see eq. [\ref{Y}]).\footnote{At early times (less than
about a day) inverse-Compton radiation is important in cooling the electrons
that are emitting synchrotron X-rays. Later the X-ray emitting
electrons are within the Klein-Nishina cutoff and are not cooled by
inverse-Compton radiation (see \citealt{FP06}).} However, one
should keep in mind that a different time dependence of $\epsilon_{\rm
X}$ (which might occur under less standard assumptions) would modify
the value of $f$ accordingly. The value of $\epsilon_{\rm obs,GRB}$
can be estimated by extrapolating the observed part of the spectrum
and modeling additional spectral components that might carry
considerable energy.

To estimate $\kappa=E_{\rm \gamma,iso}^{\rm obs}/E_{\rm
k,iso}(10\;{\rm hr})$ we calculate $E_{\rm\gamma,iso}^{\rm obs} =
f_\gamma 4\pi d_L^2(1+z)^{-1}$ directly from the observed $\gamma$-ray
fluence, $f_\gamma$, and the measured redshift, $z$.  The
denominator, $E_{\rm k,iso}(10\;{\rm hr})$, has been estimated
following \citet{FW01} and \citet{K00} \citep[see also][]{L-RZ04}
from $F_{\rm X}(10\;{\rm hr})$ using standard afterglow theory. We
reconsider this calculation in the next section.

\subsection{Estimates of the  kinetic energy from the X-ray afterglow
observations}
\label{sec:ek_est}

Using the results of \citet{GS02} we find that, if
$\nu_{\rm X}>\max(\nu_m,\nu_c)$, we can estimate the kinetic
energy from the observed X-ray flux. We use
\begin{equation}\label{E_k_iso}
E_{\rm k,iso}(t) = 9.2\times 10^{52}\frac{g(p)}{g(2.2)}(1+Y)^{4/(p+2)}
\epsilon_{e,-1}^{4(1-p)/(p+2)}\epsilon_{B,-2}^{(2-p)/(p+2)}
L_{\rm X,46}^{4/(p+2)}t_{\rm 10hr}^{(3p-2)/(p+2)}\;{\rm erg}\ ,
\end{equation}
where $L_{\rm X,iso}=L_{\rm X,46}10^{46}\;{\rm erg\;s^{-1}}$ is the
isotropic-equivalent X-ray luminosity in the range $2-10\;$keV at
a time $10\,t_{\rm 10hr}\;{\rm hr}$, both measured in the
cosmological frame of the GRB [corresponding to an observed time
of $t=10(1+z)t_{\rm 10hr}\;$hr and spectral range
$2/(1+z)-10/(1+z)\;$keV], $\epsilon_{e,-1}=\epsilon_e/0.1$,
$\epsilon_{B,-2}=\epsilon_B/0.01$, and
\begin{equation}\label{g}
g(p) = \left[\frac{(p-1)^{p-1}\exp(5.89p-12.7)}
{(5^{(p-2)/2}-1)(p-2)^{p-2}(p-0.98)}\right]^{4/(p+2)}\ .
\end{equation}
In equation (\ref{E_k_iso}), $L_{\rm X,iso}$ is evaluated through
$F_{\rm X}$ using equation (\ref{L_X}), and it is assumed that $F_{\rm
X}$ is dominated by the synchrotron component. If there is a
significant contribution to $F_{\rm X}(t_*)$ from the SSC component
then equation (\ref{E_k_iso}) would overestimate $E_{\rm k,iso}(t_*)$,
but the correct estimate could still be obtained if only the
synchrotron contribution to $F_{\rm X}(t_*)$ is
used (although in practice it might be difficult and somewhat
model-dependent to disentangle the synchrotron and SSC
components). For our fiducial parameters $(1+Y)\approx 3.7$, so
the numerical coefficient in equation (\ref{E_k_iso}) is
$3.2\times 10^{53}\;$erg for $p=2.2$ and $5.8\times
10^{53}\;$erg for $p=2.5$. This is a factor $\sim 30-60$ higher
than the numerical coefficient in equation (7) of
\citet{L-RZ04}. Part of this difference (a factor of $\sim
3-3.5$) reflects the fact that these authors did not take into
account the SSC contribution, which reduces the flux in the relevant power-law
segment of the spectrum by a factor of $(1+Y)$
\citep{SE01,GS02},\footnote{This is assuming that the SSC component
does not contribute considerably to the observed X-ray flux, which is
typically the case at $t_*=10\;$hr.}  but this does not account for
most of the discrepancy. Most of the difference is basically a
result of a higher (by a factor of $\sim 38$ for $p=2.2$) value
of $\nu_m$ that \citet{L-RZ04} use in their equation (2), given that
their expressions for $\nu_c$ and $F_{\rm\nu,max}$ (their
eqs. [3] and [4]) are very similar to those
in \citet{GS02} and that $F_{\nu>\max(\nu_m,\nu_c)} =
F_{\rm\nu,max}\nu_c^{1/2}\nu_m^{(p-1)/2}
\nu^{-p/2}\propto\nu_m^{(p-1)/2}$. Our numerical coefficient is lower
than that in equation (4) of \citet{FW01} by a factor of $\sim 3$ and
$\sim 8$ (or $\sim 12$ and $\sim 31$ if SSC is not taken into account)
for $p=2.2$ and $p=2.5$, respectively. The difference in the numerical
coefficient between \citet{L-RZ04} and \citet{FW01} is by a factor of
$\sim 100$ and $\sim 475$ for $p=2.2$ and $p=2.5$,
respectively. Using equations (\ref{epsilon_X}) and
(\ref{E_k_iso}), we can express the efficiency as
\begin{eqnarray}\label{epsilon_X2}
\epsilon_{\rm X}(t) &=& 3.5\times 10^{-3}
\left[\frac{g(p)}{g(2.2)}\right]^{-(p+2)/4}(1+Y)^{-1}
\epsilon_{e,-1}^{p-1}\epsilon_{B,-2}^{(p-2)/4}
E_{\rm k,iso,52}^{(p-2)/4}t_{\rm 10hr}^{-3(p-2)/4}
\\ \nonumber
 &=&  3.9\times 10^{-3}\,\frac{g(2.2)}{g(p)}\,(1+Y)^{-4/(p+2)}
\epsilon_{e,-1}^{4(p-1)/(p+2)}\epsilon_{B,-2}^{(p-2)/(p+2)}
L_{\rm X,46}^{(p-2)/(p+2)}t_{\rm 10hr}^{-2(p-2)/(p+2)}\ ,
\end{eqnarray}
where $E_{\rm k,iso,52}=E_{\rm k,iso}(t)/(10^{52}\;{\rm erg})$. For
our fiducial values (and for $L_{\rm X,46}$ rather than $E_{\rm
k,iso,52}$), the numerical coefficient in equation (\ref{epsilon_X2})
is $1.1\times 10^{-3}$ for $p=2.2$ and $6.3\times 10^{-4}$ for
$p=2.5$.

Equations (\ref{E_k_iso})--(\ref{epsilon_X2}) are valid for $p>2$, but
they can be easily generalized to $p\lesssim 2$ by substituting
$\epsilon_e\to\bar{\epsilon}_e(p-1)/(p-2)$, where
$\bar{\epsilon}_e=\epsilon_e\gamma_{\rm min}/\langle\gamma_e\rangle$,
$\langle\gamma_e\rangle=\int d\gamma_e(dN/d\gamma_e)\gamma_e
\left[\int d\gamma_e(dN/d\gamma_e)\right]^{-1}$ is the average
electron Lorentz factor, and the electron energy distribution is a
power law of index $p$ at low energies above $\gamma_{\rm min}$. If
the electron energy distribution is a single power law,
$dN/d\gamma_e\propto\gamma_e^{-p}$ for $\gamma_{\rm
min}<\gamma_e<\gamma_{\rm max}$, then
\begin{equation}\label{epsilon_e_ratio}
\frac{\epsilon_e}{\bar{\epsilon}_e} = \frac{\langle\gamma_e\rangle}{\gamma_{\rm
min}} = \left(\frac{p-1}{p-2}\right)\frac{1-(\gamma_{\rm
    max}/\gamma_{\rm min})^{2-p}}{1-(\gamma_{\rm
    max}/\gamma_{\rm min})^{1-p}}  = \left\{\matrix{
\approx (p-1)/(p-2) & \quad p > 2\ , \cr\cr
\ln(\gamma_{\rm max}/\gamma_{\rm min}) & \quad p=2\ , \cr\cr
\approx (\gamma_{\rm max}/\gamma_{\rm min})^{2-p}(p-1)/(2-p) & \quad 1 < p < 2\
,
\cr\cr
(\gamma_{\rm max}/\gamma_{\rm min})/\ln(\gamma_{\rm max}/\gamma_{\rm
  min}) & \quad p = 2\ , \cr\cr
\approx(\gamma_{\rm max}/\gamma_{\rm min})(1-p)/(2-p) & \quad p < 1\ .}\right.
\end{equation}
This results in slightly modified equations:
\begin{equation}\label{E_k_iso_bar}
E_{\rm k,iso}(t) = 1.19\times
10^{52}\frac{\bar{g}(p)}{\bar{g}(2.2)}(1+Y)^{4/(p+2)}
\bar{\epsilon}_{e,-1}^{4(1-p)/(p+2)}\epsilon_{B,-2}^{(2-p)/(p+2)}
L_{\rm X,46}^{4/(p+2)}t_{\rm 10hr}^{(3p-2)/(p+2)}\;{\rm erg}\ ,
\end{equation}
\begin{equation}\label{g_bar}
\bar{g}(p) = \left[\frac{(p-2)\exp(5.89p-12.7)}
{(5^{(p-2)/2}-1)(p-0.98)}\right]^{4/(p+2)}\ ,
\end{equation}
\begin{eqnarray}\label{epsilon_X2_bar}
\epsilon_{\rm X}(t) &=& 3.01\times 10^{-2}
\left[\frac{\bar{g}(p)}{\bar{g}(2.2)}\right]^{-(p+2)/4}(1+Y)^{-1}
\bar{\epsilon}_{e,-1}^{p-1}\bar{\epsilon}_{B,-2}^{(p-2)/4}
E_{\rm k,iso,52}^{(p-2)/4}t_{\rm 10hr}^{-3(p-2)/4}
\\ \nonumber
 &=&  3.03\times 10^{-3}\,\frac{\bar{g}(2.2)}{\bar{g}(p)}\,(1+Y)^{-4/(p+2)}
\bar{\epsilon}_{e,-1}^{4(p-1)/(p+2)}\epsilon_{B,-2}^{(p-2)/(p+2)}
L_{\rm X,46}^{(p-2)/(p+2)}t_{\rm 10hr}^{-2(p-2)/(p+2)}\ .
\end{eqnarray}
Note that the numerical coefficient was calculated in \citet{GS02}
only for $p=2.2,\,2.5,\,3$ and interpolated between these values.
Extrapolating that formula to $p\lesssim 2$ could potentially be
very inaccurate.

For simplicity we use the expression for $Y$ that is valid in
the fast-cooling regime,
\begin{equation}\label{Y}
Y = \frac{(1+4\epsilon_e/\epsilon_B)^{1/2}-1}{2}\approx\left\{
\matrix{(\epsilon_e/\epsilon_B)^{1/2} & \quad
  \epsilon_e/\epsilon_B\gg 1\cr\cr
\epsilon_e/\epsilon_B & \quad \epsilon_e/\epsilon_B\ll 1}\right.\ ,
\end{equation}
which is still reasonable at $10\;$hr, even if it is slightly after
the transition to slow cooling. More generally,
$\epsilon_e/\epsilon_B$ should be multiplied by $\epsilon_{\rm
rad}\approx\min\left[1,(\nu_m/\nu_c)^{(p-2)/2}\right]$, where for
$p<2$ and $\nu_c<\nu_{\rm max}$ we have $\epsilon_{\rm rad}\approx 1$.

Before applying these relations to observed bursts we remark on a
common misconception concerning the dependence of $E_{\rm k,iso}$
that is inferred from $F_{\rm X}$ on $\epsilon_B$ and
$\epsilon_e$. It has been argued that $E_{\rm k,iso}$ is very
insensitive to the exact value of $\epsilon_B$
\citep[e.g.,][]{FW01,P01}. This follows from the observation (see
eq. [\ref{E_k_iso}]) that for $\nu_{\rm X} > \max(\nu_m,\nu_c)$,
$E_{\rm k,iso}
\propto(1+Y)^{4/(p+2)}\epsilon_e^{-4(p-1)/(p+2)}\epsilon_B^{-(p-2)/(p+2)}$,
which suggests that $E_{\rm k,iso}$ depends very weakly on
$\epsilon_B$. However, this holds only in the limit where
$(1+Y)\approx 1$, which corresponds to $Y\ll 1$ and
$\epsilon_B\gg\epsilon_e$, whereas observations suggest that the
opposite limit typically applies, $\epsilon_B\ll\epsilon_e$, in
which case $(1+Y)\approx Y\approx (\epsilon_e/\epsilon_B)^{1/2}$
and $E_{\rm k,iso}\propto\epsilon_B^{-p/(p+2)}$. This is a
significantly stronger dependence on $\epsilon_B$. (Note that
the inferred value of $\epsilon_B$ varies by $\sim 2$ orders of
magnitude among
different afterglows, from $\sim 10^{-3}$ to $\sim 0.1$,
corresponding to a change of more than an order of magnitude in
the estimated value of $E_{\rm k,iso}$.) The dependence of $E_{\rm
k,iso}$ on $\epsilon_e$ is stronger, $\propto
\epsilon_e^{-4(p-1)/(p+2)}$ (i.e., slightly steeper than an inverse
linear relation) in the limit $\epsilon_B\gg\epsilon_e$, but
only $\propto\epsilon_e^{-2(2p-3)/(p+2)}$ in the more relevant
limit of $\epsilon_B\ll\epsilon_e$. In the latter case the
dependence of $E_{\rm k,iso}$ on $\epsilon_B$ is rather similar
to its dependence on $\epsilon_e$.\footnote{Quite often $Y\sim
1-2$ is inferred, in which case neither of the asymptotic limits
$Y\ll 1$ and $Y\gg 1$ is applicable and the dependence of $E_{\rm k,iso}$ on
$\epsilon_B$ and $\epsilon_e$ does not have a power-law
form but rather the more complex form given by eqs. (\ref{E_k_iso})
and (\ref{Y}).} However, $\epsilon_e$ appears to vary much less
than $\epsilon_B$ among different afterglows, only covering a range
of about one order of magnitude (between $\sim 10^{-1.5}$ and $\sim
10^{-0.5}$), which corresponds to a variation in $E_{\rm k,iso}$
by a factor of $\sim 4$. It is also worth noting that the
expression for $E_{\rm k,iso}$ has some (nontrivial) dependence
on the value of $p$ (see eq. [\ref{g}]).

\subsection{Pre-Swift GRBs}
\label{sec:pre_swift}

Table \ref{pre-Swift}\/ shows the estimated values of $\kappa$ deduced
from the observational properties of 17 pre-{\it Swift}\/ GRBs with
known redshifts, using the samples of \citet{BKF03} and
\citet{BFK03}. We provide the values $E_{\rm k,iso,10hr}=E_{\rm
k,iso}(10\;{\rm hr})$ and $\kappa$ for our fiducial parameter values
($\epsilon_e=0.1$, $\epsilon_B=0.01$, and $p=2.2$). The value of
$E_{\rm k,iso,10hr}$ and therefore of $\kappa$ depends on the values
of the microphysical parameters ($\epsilon_e$, $\epsilon_B$, and $p$)
that are not well known. Therefore we also calculate $E_{\rm
k,iso,10hr}$ and $\kappa$ (for those GRBs for which this is possible)
using the values of the microphysical parameters inferred from the
fits to the broad-band afterglow data that were performed by
\citet[][denoted by PK02, using Table 2 therein]{PK02} and by
\citet[][denoted by Y03, using Table 1 therein]{Y03}. Finally, we
compare the values we obtain for $E_{\rm k,iso,10hr}$ using equation
(\ref{E_k_iso}) to those obtained for $E_{\rm k,iso}(1\;{\rm
day})=E_{\rm k,iso,1d}$ by Y03 and those obtained for $E_{\rm k,iso}(10\;{\rm
hr})\approx 0.5E_{\rm k,iso,0}$ by PK02. Also shown in Table
\ref{pre-Swift}\/ are the corresponding values of $\kappa$, including
$\kappa_{\rm 1d}=E_{\rm\gamma,iso}/E_{\rm k,iso,1d}^{Y03}$.

When using the fiducial parameters $\epsilon_e=0.1$,
$\epsilon_B=0.01$, and $p=2.2$ our estimates for $E_{\rm k,iso,10hr}$
are significantly larger than the estimates of \citet{L-RZ04}, who use
the same values. This can be traced to the difference in
the numerical factor that appears in equation (\ref{E_k_iso}). These
relatively large values of $E_{\rm k,iso,10hr}$ lead to a typical
value of $\kappa$ around $0.1-0.2$, for which the $\gamma$-ray
efficiency problem would not be very severe. Similar results were
obtained by \citet{FP06}, whose estimates for $E_{\rm k,iso,10hr}$ are
within a factor of 2 of those presented here.

The situation is different when we use the values of the
microphysical parameters from the pre-{\it Swift}\/ afterglow
fits. In these cases the values of $E_{\rm k,iso,10hr}$ are
typically lower, resulting in higher estimates for $\kappa$. The
estimates of the kinetic energy from the PK02 fits to the
afterglow data (rather than from using eq. [\ref{E_k_iso}]) are
generally the lowest.\footnote{In PK02 the values of $E_{\rm
k,iso}(10\;{\rm hr})$ are typically a factor of $\sim 2$ smaller
than $E_{\rm k,iso,0}$ due to radiative losses at early times (A.
Panaitescu, personal communication; no energy injection is assumed
in that work).} The corresponding values of $\kappa$ are close to
unity. These results reflect the pre-{\it Swift}\/ inference that
there exists a rough equality between the isotropic-equivalent
$\gamma$-ray and late (10~hr) kinetic energies. The comparison
between the detailed calculations and those based on equation
(\ref{E_k_iso}) may suggest that we might be better off adopting
different fiducial parameters (e.g., $\epsilon_e = 0.3$,
$\epsilon_B = 0.08$, $p = 2.2$). We also note that the values of
$E_{\rm k,iso,10hr}$ obtained by substituting the values of the
microphysical parameters from the afterglow fits into equation
(\ref{E_k_iso}) are generally higher (corresponding to lower
values of $\kappa$) compared to those derived directly by those
fits. The ``typical'' values of the microphysical parameters
inferred from the fits are roughly $\epsilon_e\approx 0.3$,
$\epsilon_B\approx 0.03$, $p\approx 2.2$.

The values of $\kappa$ obtained here are crucial to the overall
estimate of $\epsilon_\gamma$ and to the origin of the ``high
efficiency problem.'' Considering Table \ref{pre-Swift}, one should
proceed with care in view of the large dispersion in the different
estimates of $E_{\rm k,iso,10hr}$. An example is a factor of 4.7
between the independent estimates of PK02 and of Y03 for GRB~000926
(and a factor of 2.5 in the opposite direction for GRB~970508). The
dispersion is even greater between the estimates of $E_{\rm k,iso}$
from the afterglow fits and the values obtained using equation
(\ref{E_k_iso}) for the same values of the microphysical parameters
--- a factor of 18 for Y03 (and 11 for PK02) for GRB~000926 .

\subsection{Swift GRBs}
\label{sec:swift}

The new result found by {\it Swift}\/ is the appearance of a rapid
decay followed by a shallow decline phase in the X-ray afterglow. For
seven out of the ten {\it Swift}\/ GRBs considered here there was a
clear observation of the two breaks in the light curve, at $t_{\rm
break,1}$ and $t_{\rm break,2}$ (the beginning and end, respectively,
of the flattish segment of the X-ray light curve).  If we interpret
the shallow decline as arising from an additional injection of energy
into the afterglow shock (see \S~\ref{sec:introduction}) then the
ratio of the X-ray fluxes at the end and at the beginning of this
phase can be used to estimate $f$ (the corresponding ratio of the
kinetic energies) for these bursts \citep{Nousek05}.  A lower limit,
$f_{\rm min}$, is obtained using the flux decrement at $t_{\rm
break,1}$ relative to the extrapolation back to that time of the
late-time ($t>t_{\rm break,2}$) flux. An approximate upper limit,
$f_{\rm max}$, is obtained by assuming that the flat part of the
emission from the forward shock starts at $t_{\rm dec}\sim T_{\rm
GRB}$ and is simply buried underneath the tail of the prompt emission
at $t<t_{\rm break,1}$. While formally $f_{\rm min} < f \lesssim
f_{\rm max}$, it is reasonable, in the context of the basic jet model,
that the assumption made to calculate $f_{\rm
max}$ is basically applicable, so that $f\sim f_{\rm max}$.
Under this assumption one infers $f\gtrsim 10$, and in some cases
even a much larger value of $f$ ($\sim 10^2-10^3$).

The results for $f$ need to be combined with an estimate of
$\kappa$. Table \ref{Swift}\/ shows the values of $E_{\rm k,iso,10hr}$
and $\kappa$ for the ten {\it Swift}\/ GRBs with known redshifts from
the \citet{Nousek05} sample, estimated using equation (\ref{E_k_iso})
with $\epsilon_e=0.1$ and $\epsilon_B=0.01$. The value of $p$ was
derived using the measured spectral slope in the X-ray band
(attempting to fit it into the range $2<p<3$ if allowed within the
measurement errors).  Lacking any broad-band fits to {\it Swift}\/
bursts, this is the best direct evidence that we have so far from
these data.  The resulting values of $\kappa$ are similar to those
from the pre-{\it Swift}\/ era (see Table \ref{pre-Swift}). With the
exception of GRB~050401, for which $\kappa=0.41$, and a few other
bursts for which we only have upper limits, the inferred values of
$\kappa$ are less than 0.1. If this is the correct value of $\kappa$
then, using
\begin{equation}\label{eps_g_fk1}
\epsilon_\gamma = \left(1+\frac{\epsilon_{\rm obs,GRB}}{\kappa
  f}\right)^{-1}
\end{equation}
(see eq.~[\ref{obs_ratio}]), we find that with $f \sim 10$ the overall
$\gamma$-ray efficiency is not larger than $\sim 0.5$ (assuming
$\epsilon_{\rm obs,GRB}\sim 1$). A similar conclusion was reached by
\citet{FP06}.

One may question this conclusion in view of the fact that, in pre-{\it
Swift}\/ bursts, broad-band analyses of the afterglow data generally
resulted in a significantly lower values of $E_{\rm k,iso,10hr}$, and
correspondingly higher values of $\kappa$, compared to those obtained
from equation (\ref{E_k_iso}) with the same fiducial values of the
microphysical parameters ($\epsilon_e=0.1$ and $\epsilon_B=0.01$; see
Table \ref{pre-Swift}). Furthermore, the choice of the fiducial values
of the microphysical parameters is somewhat arbitrary, and it affects
the resulting values of $E_{\rm k,iso,10hr}$ and $\kappa$.  It is
reasonable to expect that the values of the microphysical parameters
that would have been inferred from a broad-band fit to the afterglow
data of the {\it Swift}\/ bursts would have led to higher estimates of
$E_{\rm k,iso,10hr}$ and $\kappa$ that were similar to those derived
for the pre-{\it Swift}\/ GRBs. The latter values, however, vary
among the different estimates, from as high as $\sim 5-8$ to as low
as $\sim 0.1-0.3$ (see Table \ref{pre-Swift}).  In light of this, one
may adopt a ``typical'' value of $\kappa\sim 1$, keeping in mind that
there is an uncertainty of almost an order of magnitude in this value.

The adoption of this higher value of $\kappa$ ($\sim 1$) for the {\it
Swift}\/ GRBs, similar to the values inferred from broad-band modeling
of pre-{\it Swift}\/ sources, together with the interpretation of the
shallow decay phase as arising from energy injection (and hence $f
\sim 10$) leads to the conclusion that typically $\epsilon_\gamma \sim
0.9$, and in some cases $\epsilon_\gamma$ is even as high as $\sim
99\%$ (or, equivalently, $1-\epsilon_\gamma\approx\epsilon_{\rm
obs,GRB}/\kappa f$ is as low as $\sim 10^{-3}-10^{-2}$). { Such a high
$\gamma$-ray efficiency would be extremely hard to produce in the
internal shocks model (see \S~\ref{sec:introduction}).

If, on the other hand, $\kappa\sim 0.1$ and there is energy injection
(i.e. $f\sim 10$), or if $\kappa\sim 1$ and there is no energy
injection (i.e. $f=1$; see \S~\ref{sec:epsilon_X}), then this would
imply a significantly lower typical $\gamma$-ray efficiency,
$\epsilon_\gamma\sim 0.5$, although in some cases $\epsilon_\gamma$
would still be as high as $\sim 90\%$ (or, equivalently,
$1-\epsilon_\gamma\approx\epsilon_{\rm obs,GRB}/\kappa f$ would still
be as low as $\sim 10^{-2}-10^{-1}$).  Even the latter, more moderate,
requirements on the $\gamma$-ray efficiency are not easily satisfied
in the internal-shocks model, athough they might possibly still be
accommodated in this scenario \citep{KPS97,K99,GSW01}.  Finally, if
$\kappa\sim 0.1$ and the shallow decline does not arise from energy
injection (i.e., $f=1$) but, say, from varying afterglow efficiency
(as discussed in \S~\ref{sec:epsilon_X}), then the $\gamma$-ray
efficiency would typically be $\epsilon_\gamma \sim 0.1$. In the
latter case the results are fully consistent with the predictions of the
internal-shocks model.

\subsection{Some Caveats}
\label{sec:assess}

The discussion so far was relevant to the power-law segment of the
spectrum where $\nu_{\rm X}>\max(\nu_m,\nu_c)$.  If, instead,
$\nu_m<\nu_{\rm X}<\nu_c$, we have only a lower limit on the
value of $E_{\rm k,iso}$ from this consideration (i.e., from eq.
[\ref{E_k_iso}]) that corresponds to an upper limit on the value of
$\kappa$, the true value being smaller than this upper bound by a
factor $[\nu_c(10\;{\rm hr})/\nu_{\rm X}]^{2/(p+2)}$ (which, however, is
not typically expected to be very large).

It is possible that there is a contribution to $F_{\rm X}(t_{\rm
dec})$ from a component of the outflow that is not along the line of
sight, or for some other reason did not contribute to the observed
$\gamma$-ray emission. In this case equations (\ref{obs_ratio}) and
(\ref{E_k_iso}) would provide a lower limit on $\epsilon_\gamma$,
rather than directly determine its value. If both the prompt
$\gamma$-ray emission and $F_{\rm X}(t_{\rm dec})$ were dominated by
emission from angles $\theta>1/\Gamma$ relative to the line of
sight (where $\Gamma$ is the Lorentz factor of the outflow),
then again equations (\ref{obs_ratio}) and (\ref{E_k_iso}) would
provide a lower limit on $\epsilon_\gamma$, since the beaming of
radiation away from the line of sight is expected to be either
comparable or somewhat smaller during the afterglow emission at $t_{\rm
dec}$ compared to the prompt GRB.

It is also possible that some fraction $\epsilon_*$ of the original
energy $E_{\rm iso,0}$ ended up in a totally different form, i.e., was
not radiated during the prompt emission and did not end up in the
kinetic energy of the afterglow shock. This could occur, for example,
if along some directions a forward shock is not formed (or at least
not formed efficiently) for a very pure Poynting-flux outflow.  In
such a case some of the energy (potentially even most of the energy)
could be carried out to very large distances (in principle out to
infinity) in electromagnetic form (such as low-frequency
electromagnetic waves).\footnote{Both the formation of a forward shock
and the ability of energy to escape to infinity in electromagnetic
form have not yet been fully worked out \citep[e.g.,][]{MM96,SU00}.}
Alternatively, a good fraction of the energy might be carried away in
high-energy cosmic rays and neutrinos \citep{Waxman95,WB97} and thus
would not contribute to the kinetic energy of the afterglow shock. In
this case we have $E_{\rm k,iso,0} = (1-\epsilon_\gamma-\epsilon_*)
E_{\rm iso,0}$ and we need to make the substitution
$(1-\epsilon_\gamma) \to (1-\epsilon_\gamma-\epsilon_*)$ everywhere,
so the estimate (\ref{eps_g_fk1}) for $\epsilon_\gamma$ (assuming
$\epsilon_*=0$) should be multiplied by $(1-\epsilon_*)$,
\begin{equation}\label{eps_g_fk*}
\epsilon_\gamma = (1-\epsilon_*)\left(1+\frac{\epsilon_{\rm
  obs,GRB}}{\kappa f}\right)^{-1}\ .
\end{equation}
This means that $\epsilon_\gamma<1-\epsilon_*$ (even for $\kappa
f\gg\epsilon_{\rm obs,GRB}$), and therefore $\epsilon_*\gtrsim 0.5$
would imply $\epsilon_\gamma\lesssim 0.5$ . Thus, even when one
infers $\epsilon_\gamma/(1-\epsilon_\gamma)\gg 1$ and hence
$1-\epsilon_\gamma\ll 1$ under the usual assumption that
$\epsilon_*=0$ (or at least $\epsilon_*\ll 1-\epsilon_\gamma$), the
intrinsic $\gamma$-ray efficiency might still be significantly
smaller, and is $\epsilon_\gamma\lesssim 0.5$ for $\epsilon_*\gtrsim
0.5$. Note that, in order for GRBs to produce the highest-energy cosmic
rays, their energy should be comparable to that of the prompt
$\gamma$-rays \citep{Waxman95,Waxman04},
i.e., $\epsilon_\gamma\lesssim\epsilon_*$ (the inequality arising since
there might be other forms of energy that escape the prompt emission
site without being directly detected), and therefore
$\epsilon_\gamma\lesssim 0.5$.

One should, however, keep in mind that high-energy cosmic rays
and neutrinos must tap the same dissipated energy that also
powers the prompt $\gamma$-ray emission.  Therefore, for the
same observed energy in $\gamma$-rays and inferred kinetic
energy in the afterglow, in addition to increasing the required
total initial energy $E_{\rm iso,0}$ by a factor of
$(1-\epsilon_*)^{-1}$, these particles would also increase the
required dissipated energy $\epsilon_{\rm dis,GRB}E_{\rm iso,0}$
and the fraction $\epsilon_{\rm dis,GRB}$ of the dissipated
energy that ends up in $\gamma$-rays.\footnote{In the case of a
very pure Poynting flux, the escaping energy $E_{*,{\rm iso}}$
is the fraction that did not dissipate. Therefore, while $E_{\rm
iso,0}$ increases by a factor of $(1-\epsilon_*)^{-1}$, the
dissipated energy $\epsilon_{\rm dis,GRB}E_{\rm iso,0}$ remains
unchanged (assuming other efficiencies do not change), and thus
$\epsilon_{\rm dis,GRB}$ decreases by a factor of
$(1-\epsilon_*)^{-1}$.}  In other words, $\epsilon_{\rm dis,GRB}
\geq\epsilon_\gamma+\epsilon_*=(E_{\rm\gamma,iso}+E_{*,{\rm
iso}})/(E_{\rm k,iso,0}+E_{\rm\gamma,iso}+E_{*,{\rm
iso}})>E_{\rm\gamma,iso}/(E_{\rm k,iso,0}+E_{\rm\gamma,iso})$,
where $E_{*,{\rm iso}}=\epsilon_*E_{\rm iso,0}$ is the
(isotropic-equivalent) energy in cosmic rays and neutrinos, and
$E_{\rm k,iso,0}$ and $E_{\rm\gamma,iso}$ are determined (at
least in principle) by observations.  Clearly, $E_{*,{\rm iso}}$
increases the lower limit on $\epsilon_{\rm
dis,GRB}$. Nevertheless, the fact that $E_{*,{\rm iso}}$ may
reduce $\epsilon_\gamma$ to $\lesssim 0.5$ even for $\kappa f\gg
1$ makes it possible to have $\epsilon_{e,{\rm GRB}}\lesssim
0.5$, which should be easier to accommodate for shock
acceleration in a proton--electron plasma. Still, dissipating
and getting rid of almost all of the energy (through radiation,
cosmic rays, neutrinos, etc.)  and leaving only a small fraction
of the original energy in the kinetic energy of the forward
shock, as is required for $\kappa f\gg 1$, is not an easy task
for any model of the prompt emission.

\section{Efficiency of the X-Ray Afterglow Emission}
\label{sec:epsilon_X}

As noted in \S~\ref{sec:introduction}, one of the new features
discovered by {\it Swift}\/ is the early shallow decline phase: $F_{\rm
X} \propto t^{-\alpha}$ with $0.2 \lesssim \alpha \lesssim 0.8$ for
$t_{\rm break,1} \lesssim t \lesssim t_{\rm break,2}$. During this
phase $t F_{\rm X}(t)$ increases with time. Using the definition of
$\epsilon_{\rm X}(t)$ (eq. [\ref{epsilon_X}]) and the relation between
$L_{\rm X,iso}$ and $F_{\rm X}$ (eq. [\ref{L_X}]), we find that
\begin{equation}\label{ratio1} \frac{\epsilon_{\rm X}(t)E_{\rm
k,iso}(t)}{tF_{\rm X}(t)} = 4\pi d_L^2(1+z)^{\beta-\alpha-1}\
\end{equation} is constant in time. Note that if $\nu$ and $t$ in the
expression for $F_{\rm X}(t)$ were referred to the GRB rest frame rather
than to the observer frame then the factor $(1+z)^{\beta-\alpha-1}$ on
the right-hand side of equation (\ref{ratio1}) would be eliminated, and with
it any potential (weak) time dependence resulting from a possible
temporal variation of $\alpha$ or $\beta$.

Now, if the observed frequencies satisfy $\nu_{\rm
X}>\max(\nu_m,\nu_c)$, and $p>2$, then equation
(\ref{epsilon_X_standard}) is applicable. Furthermore, if the
afterglow shock evolves according to the adiabatic self-similar
solution of \citet{BM76} and the fractions of the postshock internal
energy in electrons ($\epsilon_e$) and in magnetic field
($\epsilon_B$) are constant in time, then $\nu_m\propto t^{-3/2}$.
Under these circumstances $\epsilon_{\rm X}\propto t^{-3(p-2)/4}$
decreases (slowly) with time. As can be seen from equation
(\ref{ratio1}), $\epsilon_{\rm X}(t)E_{\rm k,iso}(t)\propto tF_{\rm
X}(t)$.  Therefore, the observed rise in $tF_{\rm X}(t)$ implies a
similar rise in $\epsilon_{\rm X}(t)E_{\rm k,iso}(t)$. Given the
expected decrease of $\epsilon_X(t)$ with $t$ for $p>2$, the slowly
decaying portion of the light curve has been attributed by several
researchers to an increase in $E_{\rm k,iso}(t)$, i.e., to some sort
of energy injection into the forward shock
\citep[e.g.,][]{Nousek05,Panaitescu05,Zhang05,GK06}.

It is, however, conceivable that the rise in $tF_{\rm X}(t)$
corresponds, at least in part, to an increase of $\epsilon_{\rm X}(t)$
with time. One way in which this could be brought about is if $p$ were
$<2$ (assuming $\epsilon_e$ and $\epsilon_B$ remain
constant).\footnote{It is in principle possible that $\epsilon_{\rm
X}$ could increase with time on account of its dependence on $Y$ (see
eq. [\ref{epsilon_X_standard}) even if $\epsilon_e$ and $\epsilon_B$
remained constant and $p$ were $>2$, given that $Y$ decreases with
time in the slow-cooling regime ($\nu_m<\nu_c$) that is relevant for
$p>2$. However, one can show that, to be relevant during the
early-afterglow phase, this would require unrealistically high values
of $p$.} In this case $N(\gamma_e)\propto\gamma_e^{-p}$ for
$\gamma_{\rm min}<\gamma_e<\gamma_{\rm max}$ and $\epsilon_{\rm
X}\sim(1+Y)^{-1}\epsilon_e(\nu_{\rm max}/\nu_{\rm X})^{(p-2)/2}$ [the
same as eq. (\ref{epsilon_X_standard}) for $p<2$, but with $\nu_{\rm
max}$ replacing $\nu_m$], where $\nu_{\rm max}=\nu_{\rm
syn}(\gamma_{\rm max})\propto\gamma B\gamma_{\rm
max}^2\propto\gamma^4\rho_{\rm ext}^{1/2}\ \propto t^{-3/2}$ [where
$B$ is the comoving magnetic field amplitude and $\rho_{\rm ext}$ is
the external density; the same as the scaling of $\nu_m=\nu_{\rm
syn}(\gamma_{\rm min})$ for $p>2$], so $\epsilon_{\rm X}\propto
t^{3(2-p)/4}$. This time dependence is the same as for $p>2$, but in
this case $\epsilon_{\rm X}$ increases with time whereas for $p>2$ it
decreases. Similarly, $F_{\rm X}\propto t^{-(3p-2)/4}$, just as for
$p>2$, except that for $p<2$ this corresponds to a decay rate flatter
than $t^{-1}$. One possible difference between the two cases is that
for $p<2$ and $\nu_c<\nu_{\rm max}$ radiative losses are not always
negligible since most of the energy in the electrons is at
$\gamma_e\sim\gamma_{\rm max}$ and is therefore radiated away.  Thus,
unless $\epsilon_e\ll 1$, radiative losses could be significant and
would tend to steepen the flux decay rate and make it harder to
achieve a flattish light curve. We also note that the X-ray spectral
slope for $p<2$ [assuming $\nu_{\rm X}>\max(\nu_c,\nu_m)$] is
$\beta_{\rm X}=p/2<1$, which in many cases is inconsistent with the
observed value \citep{Nousek05}, so this explanation of the shallow
decay of $F_{\rm X}$ might only apply to a subset of the sources
\citep[see Fig. 8 of][]{Nousek05}.

An alternative possibility for $\epsilon_{\rm X}(t)$ to increase
with $t$ is for either one (or both) of the microphysical
parameters $\epsilon_e$ and $\epsilon_B$ to increase with time.
Using the dependence of $\nu_m$ on these parameters
\citep[e.g.,][]{SPN98}, we find that, for $p>2$, $\epsilon_{\rm X}
\propto \epsilon_e^{(p-1)} \epsilon_B^{(p-2)/4}$ when $\epsilon_e
\ll \epsilon_B$ and $\epsilon_{\rm X} \propto \epsilon_e^{(p-3/2)}
\epsilon_B^{p/4}$ when $\epsilon_e \gg \epsilon_B$. This applies
when parameterizing in terms of $E_{\rm k,iso}$ (which is not
measured directly), whereas a parameterization in terms of $L_{\rm
X,iso}$ (which is measured directly) yields $\epsilon_{\rm X}
\propto \epsilon_e^{4(p-1)/(p+2)} \epsilon_B^{(p-2)/(p+2)}$ when
$\epsilon_e \ll \epsilon_B$ and $\epsilon_{\rm X} \propto
\epsilon_e^{2(2p-3)/(p+2)} \epsilon_B^{p/(p+2)}$ when $\epsilon_e
\gg \epsilon_B$ (see eq. [\ref{epsilon_X}]). Table 3 of
\citet{Nousek05} provides the values of $\Delta\alpha$ --- the
change in the temporal decay index across the break in the light
curve at $t_{\rm break,2}$ --- for nine {\it Swift}\/ GRBs in
which it could be measured reliably, and shows that typically
$0.5\lesssim\Delta\alpha\lesssim 1$. In our context, if
$\epsilon_e\propto t^{\alpha_e}$ and $\epsilon_B\propto
t^{\alpha_B}$ at $t<t_{\rm break,2}$, then attributing the
flattish decay phase to a growth of $\epsilon_e$ and/or
$\epsilon_B$ with time requires (in the limit
$\epsilon_e\gg\epsilon_B$) that $\Delta\alpha =
2(2p-3)/(p+2)\alpha_e + \alpha_Bp/(p+2)$.  For $2 < p < 3$, $0.5 <
p/(p+2) < 0.6$ and $0.5 < 2(2p-3)/(p+2) < 1.2$. Therefore, for
$p\approx 2$, $\alpha_e+\alpha_B\approx 2\Delta\alpha\sim 1-2$ and
for $\Delta\alpha\sim 0.5$, a roughly linear growth with time of
either $\epsilon_e$ or $\epsilon_B$ (or of their product) is
required. For $p \sim 2.6$ and $\Delta\alpha \sim 1$, a linear
growth of $\epsilon_e$ and a constant $\epsilon_B$ would work
(i.e. $\alpha_e \approx 1$ and $\alpha_B = 0$). If $Y\sim 1$
(rather than $Y\gg 1$ or $Y\ll 1$), the dependence of
$\epsilon_{\rm X}$ on $\epsilon_e$ and $\epsilon_B$ is no longer a
simple power law, requiring a similarly nontrivial dependence of
$\epsilon_e$ and/or $\epsilon_B$ on the observed time $t$ (insofar
as $F_{\rm X}$ is indeed a pure power law in $t$ during the
flattish decay phase). A physical scenario will need to account
both for this behavior and for why the time evolution of the
microphysical parameters effectively terminates at $t_{\rm
break,2}$.

The magnetic-energy parameter $\epsilon_B$ could reflect either the
structure of the ambient magnetic field or postshock
field-amplification processes. In the former case an increase of
$\epsilon_B$ with time could be caused by an increase of the
magnetization parameter $\sigma = B_{\rm ext}^2/4\pi \rho_{\rm
ext}c^2$ of the ambient gas with distance from the source, which might
occur in certain GRB progenitor models \citep[e.g.,][]{KG02}. In the
latter case one cannot at present identify a natural reason for
$\epsilon_B$ to increase during the early afterglow phase, but future
theoretical advances \citep[see, e.g.,][]{Medvedev05} might make it
possible to study the evolution of shock-generated magnetic fields
over time scales that are long enough to address this question. The
value of $\epsilon_e$ might also be linked to the changing shock
parameters (in particular, the shock Lorentz factor). However, in this
case, again, our current level of understanding does not allow us to
make a specific prediction.

The possible increase of the afterglow radiation efficiency with
time during the early phases of the X-ray light curve may also
help to lower the estimate of the $\gamma$-ray radiative
efficiency and thereby alleviate the constraints on the
internal-shocks model. If at early times ($t<t_*$)
$\epsilon_{\rm X}$ increased with $t$, then (by
eq. [\ref{ratio1}]) $E_{\rm k,iso,0}$ would be underestimated,
and therefore the parameter $f$ and the value of
$\epsilon_\gamma$ would be overestimated. The prompt emission
from internal shocks could in principle be observable even if
the very early radiation from the external (forward) shock is
weak because of a low value of $\epsilon_B$ or of a possible
suppression of $\epsilon_e$ when the shock Lorentz factor is
high. This is because the value of $\epsilon_B$ in the internal
shocks might be determined by a comparatively strong magnetic
field advected from the central source \citep[e.g.,][]{SDD01}
and because (in contrast to the forward shock at the
deceleration time, whose Lorentz factor is $\gtrsim 10^2$) the
internal shocks are only mildly relativistic.  If the afterglow
emission efficiency recovers its commonly assumed value
(equation [\ref{epsilon_X_standard}]) at $t\gtrsim t_{\rm
break,2}$ then one could in principle have $f \sim 1$,
$\kappa\sim 1$, and $\epsilon_\gamma\sim 0.5$. In this case
$E_{\rm k,iso}$ remains constant while the flattish flux decay
is caused by an increase of $\epsilon_X$ with time.  However, if
even at late times the afterglow efficiency is only a fraction
$\delta\ll 1$ of its standard value [where $\delta$ might, for
example, correspond to the fraction $\xi_e$ of the electrons
behind the forward shock that are accelerated to relativistic
energies, which could conceivably be $\ll 1$ \citep{EW05}] then
$\kappa\sim\delta\kappa_{\rm standard}\sim\delta\ll 1$ and
$\epsilon_\gamma\sim \delta$ (for $\epsilon_{\rm obs,GRB}\sim
1$). In this case $E_{\rm k,iso}$ is again constant with time
and the flattish decay phase is due to $\epsilon_X$ increasing
with time; however, the value of $\epsilon_X$ at late times is
$\delta$ times its standard value (given by equation
[\ref{epsilon_X_standard}]) and the true kinetic energy in the
afterglow shock is a factor $\delta^{-1}\gg 1$ larger than the
usual estimate of $\sim 10^{51}\;$erg.

An alternative to the explanation of the flattish segment of the light
curve in terms of a prolonged energy injection into the forward shock
of a basic jet model is the possibility (in a generalized jet model)
of distinct {\em spatial} components, some of which only contribute to
the afterglow emission at later times. This situation could arise if
(1) some of the ejecta have lower initial Lorentz factors that result
in longer deceleration times, so only a small fraction of the injected
energy is transferred to the shocked external medium early on, or if
(2) radiation from components that do not move along our line of sight
is strongly beamed away from us at early times, becoming visible only
later on when deceleration causes the respective beaming cones to
widen. Case (1) is exemplified by the two-component jet model
considered in the next section, whereas case (2) might be realized in
the ``patchy shell'' model of GRB sources \citep{KP00b} and in the
``anisotropic afterglow efficiency'' scenario outlined by
\citet{EG05}.\footnote{Case (2) might also be relevant to the
two-component jet model if the line of sight to the observer lies
within the solid angle subtended by the narrow component but outside
the inner edge of the wide component.}

\section{The Two-Component Jet Model: A Case Study}
\label{sec:2comp}

As pointed out in \S~\ref{sec:introduction}, the two-component jet model
was originally invoked by \citet{PKG05} as a possible way of alleviating
the pre-{\it Swift}\/ constraints on the $\gamma$-ray emission
efficiency. In this section we interpret the {\it Swift}\/ results in the
context of this scenario, using again the parameters $\kappa$ and $f$
introduced in \S~\ref{sec:efficiency} and affirming some of the
conclusions reached in that section. In this discussion we assume that
the standard afterglow theory applies and that the microphysical
parameters do not change with time. This formulation is used to
demonstrate that the recent observations are inconsistent with parameter
values of the two-component jet model that could lead to a lower
inferred magnitude for $\epsilon_\gamma$, reinforcing the result that,
in the context of the standard afterglow theory, the {\it Swift}
measurements have tightened the constraints on the prompt-emission
efficiency. We go on to show, however, that the two-component model can
nevertheless account for the X-ray afterglow light curve of GRB sources,
including the flattish early-time segment.

The generic two-component jet model consists of a narrow and initially
highly relativistic (conical) outflow from which the prompt emission
originates, and a moderately relativistic flow that decelerates at a
significantly later time and occupies a wider (coaxial) cone. The
narrow component is the source of the prompt emission (which is
observed as a GRB if the observer's line of sight lies within, or very
close to, its opening solid angle), whereas the wide component only
makes a contribution to the afterglow emission (which becomes
observable after it decelerates). Letting $E_i$, $E_{{\rm k},i}$,
$\theta_i$, and $\eta_i$ stand for the total initial energy, initial
kinetic energy, opening half-angle, and Lorentz factor of the two
components (with $i ={\rm n,w}$ corresponding to {\em narrow} and {\em
wide}, respectively), we have
\begin{equation}
E_{\rm\gamma,iso} = \epsilon_\gamma E_{\rm n,iso} = \epsilon_\gamma
E_{\rm n}\frac{2}{\theta_{\rm n}^2}\ .
\end{equation}
The parameter $\kappa$ defined in equation (\ref{obs_ratio}) can be
expressed in this context by the relation
\begin{equation}
E_{\rm\gamma,iso}^{\rm obs}\frac{\theta_{\rm w}^2}{2} = \epsilon_{\rm
obs,GRB}\epsilon_\gamma E_{\rm n} \left({\theta_{\rm w}\over
\theta_{\rm n}}\right)^2 \approx \kappa\left(E_{\rm k,n} +E_{\rm
k,w}\right)\ , \label{kappa}
\end{equation}
where $E_{\rm k,n} = (1-\epsilon_\gamma)E_{\rm n}$ and $E_{\rm k,w}
\approx E_{\rm w}$.  Equation (\ref{kappa}) incorporates the fact that
the kinetic energy responsible for the late afterglow emission is
determined empirically by assuming a jet of half-opening angle
$\theta_{\rm w}$, or equivalently that the isotropic-equivalent
kinetic energy inferred from late time afterglow corresponds
approximately to the total kinetic energy over the fraction of the
total solid angle occupied by the wide component.

The energy that determines the early afterglow phase is
\begin{equation}
E_{\rm early,iso}= E_{\rm k,n}\frac{2}{\theta_{\rm
n}^2} = E_{\rm k,n,iso}\ ,
\end{equation}
whereas the late afterglow phase is determined by
\begin{equation}
E_{\rm late,iso}= \left(E_{\rm k,n} + E_{\rm
k,w}\right)\frac{2}{\theta_{\rm w}^2}\ .
\end{equation}
The evidence from the {\it Swift}\/ observations that we do not see the
early afterglow emission above the rapidly decaying tail of the prompt
emission, and that even when it shows up it is rather weak, implies
that the early isotropic kinetic energy $E_{\rm early,iso}$ is a
factor $f \sim 10$ smaller than the late isotropic kinetic energy
$E_{\rm late,iso}$, where we reintroduced the parameter $f$ defined in
equation (\ref{obs_ratio}). This implies that
\begin{equation}\label{ratio}
{E_{\rm k,n} \over \theta_{\rm n}^2} = \frac{E_{\rm k,n} +E_{\rm
k,w}}{f\theta_{\rm w}^2} \approx \frac{E_{\rm k,w}}{f\theta_{\rm w}^2}\ ,
\end{equation}
which in turn implies that
\begin{equation}
E_{\rm k,n} < E_{\rm k,w}\ .
\end{equation}
Hence we can delete the term involving $E_{\rm k,n}$ from the
right-hand side of equation (\ref{kappa}) and obtain
\begin{equation}\label{Ew}
\frac{\epsilon_\gamma \epsilon_{\rm obs,GRB}}{(1-\epsilon_\gamma)} =
\kappa\frac{E_{\rm k,w}}{E_{\rm k,n}}\left(\frac{\theta_{\rm
n}}{\theta_{\rm w}}\right)^2 =
\kappa\frac{E_{\rm k,w,iso}}{E_{\rm k,n,iso}}\ .
\end{equation}
Expressing now $E_{\rm k,w}$ in terms of $E_{\rm k,n}$ using this last
equation and substituting into equation (\ref{ratio}) (which becomes
simply $f\approx E_{\rm k,w,iso}/E_{\rm k,n,iso}$ when one neglects
$E_{\rm k,n}$ on the r.h.s.), we rediscover the first relation in
equation (\ref{obs_ratio}), which can be expressed in the form of
equation (\ref{eps_g_fk1}) to yield $\epsilon_\gamma \sim 0.9$ for $f
\sim 10$ and $(\kappa / \epsilon_{\rm obs,GRB}) \sim 1$.

Equation (\ref{Ew}) with $\epsilon_\gamma\epsilon_{\rm obs,GRB}\approx
1$ implies that $\kappa E_{\rm k,w}/ \theta_{\rm w}^2 \approx E_{\rm
n}/ \theta_{\rm n}^2$, and hence, given that $E_{\rm k,w} \approx
E_{\rm w}$ and taking $\kappa$ to be $\sim 1$, that the narrow and
wide jet components initially have comparable isotropic-equivalent
energies. The ratio of true energies of the two components is
initially $E_{\rm w}/E_{\rm n}\approx\kappa^{-1}(\theta_{\rm
w}/\theta_{\rm n})^2\sim (4-9)\kappa^{-1}$ for reasonable ratios of
the opening half-angles. This ratio further increases by a factor
$(1-\epsilon_\gamma)^{-1}\approx \kappa f$ (see eq. [\ref{eps_g_fk1}])
during the prompt-emission phase.

As argued by \citet{PKG05}, this model could reduce the inferred value
of $\epsilon_\gamma$ if $E_{\rm k,w}/E_{\rm k,n}>1$ {\em and} $E_{\rm
k,w,iso}/E_{\rm k,n,iso} < 1$. However, the {\it Swift}\/ results, as
expressed by equation (\ref{ratio}), demonstrate that the latter ratio
is equal to $f \sim 10$, and hence that this possibility is not realized
in practice. \citet{PKG05} also discussed the ability of this scenario
to account for certain ``bumps'' in the late-afterglow light curve of
several pre-{\it Swift}\/ GRB sources. We now show that this model can
similarly account for the early-time flattening of the X-ray light
curve. The required model parameters are, however, distinctly different
from those considered by \citet{PKG05}; in fact, the fits that we
obtain reinforce the conclusion that $E_{\rm k,w,iso}/E_{\rm
k,n,iso}$ ($\approx f$) must be $\gg 1$.

The Lorentz factor of the narrow jet component does not directly enter
into our modeling of the light curve, although the usual ``compactness''
arguments for the prompt emission  \citep[e.g.,][]{LS01} imply that its
value is $\eta_{\rm n} \gtrsim 10^2$. Our interpretation of the flattish
segment of the light curve in the context of this scenario is that it
largely corresponds to the emergence of the wide component around its
deceleration time $t_{\rm dec,w} \propto (E_{\rm k,w,iso}/n_{\rm
ext})^{1/3} \eta_{\rm w}^{-8/3}$, where $n_{\rm ext}$ is the particle
density of the ambient medium at the deceleration radius. Owing to the
strong dependence of $t_{\rm dec}$ on $\eta$, we can constrain the value
of $\eta_{\rm w}$ within a factor of 2 or so.

Figure \ref{2comp} shows a tentative fit to the X-ray light curve of
GRB~050315 with the two-component jet model, whereas Figure
\ref{2comp_diff} demonstrates the effect of modifying the model
parameters. The extended flat segment of the light curve together with
the requirement that the contribution from the narrow component at
$t/(1+z)\sim 200\;$s does not overproduce the observed flux imply
$f\approx E_{\rm k,w,iso}/E_{\rm k,n,iso}\gtrsim 30$. This suggests that
$f\sim f_{\rm max}\sim 30$ (see Table \ref{Swift}) and that both $f_{\rm
min}$ and $f_{\rm max}$ might have been somewhat underestimated for this
event (since the fit to the data should produce $f\gtrsim f_{\rm min}$,
suggesting that $f_{\rm min}\sim 30$ for GRB~050315, which is higher
than the value of $f_{\rm min}=11$ derived by \citealt{Nousek05} and
shown in Table \ref{Swift}).  For our fiducial parameter values
($\epsilon_e=0.1$, $\epsilon_B=0.01$, and $p\approx 2.0-2.1$; see Table
\ref{Swift}\/ and Fig.~\ref{2comp}) we find (using eq. [\ref{E_k_iso}])
$\kappa\approx 0.034$ (see Table \ref{Swift}). For this value of
$\kappa$ the product $\kappa f$ is $\sim 1$, which implies (from
eq. [\ref{eps_g_fk1}] with $\epsilon_{\rm obs,GRB}\sim 1$) that
$\epsilon_\gamma\sim 0.5$.  The fit to the data shown in Figure
\ref{2comp} incorporates an SSC component whose contribution in this
case turns out to be comparable to that of the synchrotron component at
$t_*=10\;$hr, resulting in a decrease by a factor of $\sim 2$ in the
estimate of $E_{\rm k,iso}(10\;{\rm hr})$ (and in a corresponding
increase in the estimate of $\kappa$) in comparison with the values
inferred by using equation (\ref{E_k_iso}) (which only takes account of
the synchrotron contribution). The actual numerical difference between
these two estimates is, in fact, slightly larger (a factor $\sim 2.5$),
reflecting the fact that the fit employs a numerical scheme that is not
identical to equation (\ref{E_k_iso}). All in all, the model fit shown
in Figure \ref{2comp} yields $f\approx E_{\rm k,w,iso}/E_{\rm
k,n,iso}=30$, $\kappa\approx E_{\rm\gamma,iso}^{\rm obs}/E_{\rm
k,w,iso}\approx 0.086$, and $\kappa f\approx E_{\rm\gamma,iso}^{\rm
obs}/E_{\rm k,n,iso}\approx 2.6$, which implies $\epsilon_\gamma\approx
0.72$ for $\epsilon_{\rm obs,GRB}=1$.

One realization of a two-component relativistic outflow of the
type considered here is an initially neutron-rich,
hydromagnetically accelerated jet \citep{VPK03}. In this picture
the narrow component consists of the originally injected protons
that are gradually accelerated to $\eta_{\rm n}$; the injected
neutrons decouple from the original proton component when the jet
Lorentz factor reaches $\eta_{\rm w}$ and eventually decay to form
a distinct (wide) proton component.  In the illustrative solution
presented in \citet{VPK03}, $\eta_w \approx 15$. The {\it dashed
curve} in Figure \ref{2comp_diff} demonstrates that, to be
consistent with the data, this value of $\eta_{\rm w}$ requires a
very high external density, $n_{\rm ext} \sim 10^{3.5}\;{\rm
cm^{-3}}$ (although even for this value the fit is not as good as
for the parameters adopted in Fig.~\ref{2comp}). This can be
understood from the parameter dependence of $t_{\rm dec,w}/(1+z)$,
for which the model fit implies a value of $\sim 2\times 10^3\;$s.
To reproduce the observed flux, the value of $E_{\rm k,w,iso}$
cannot be too low. In fact, in the model fit represented by the
dashed curve we have adopted the lowest possible value of this
quantity, corresponding to an equipartition between the electron
and magnetic field energy densities. With the values of $t_{\rm
dec,w}$ and $E_{\rm k,w,iso}$ thus fixed, the inferred external
density becomes very sensitive to the value of the Lorentz factor
($n_{\rm ext} \propto \eta_{\rm w}^{-8}$).  Intermediate options,
with a somewhat higher Lorentz factor ($\eta_{\rm w} = 21$) are
also shown in Figure \ref{2comp_diff}, both for the
uniform-density case used in the previous fits ({\em dash-dotted
curve}) and for an $r^{-1.5}$ density profile (where $r$ is the
distance to the source; {\em dotted curve}). We interpret the
break in the light curve of GRB~050315 at $t/(1+z)\sim 1\;$day as
the jet break time of the wide component, $t_{\rm jet,w}/(1+z)$.
Since the jet break time is particularly sensitive to the value of
the opening half-angle, $t_{\rm jet,w}\propto(E_{\rm
k,w,iso}/n_{\rm ext})^{1/3}\theta_{\rm w}^{8/3}$, this allows us
to constrain the value of $\theta_{\rm w}$. Given that $t_{\rm
jet,w}/t_{\rm dec,w}\propto (\eta_{\rm w}\theta_{\rm,w})^{8/3}$,
it is seen that any reduction in the fitted value of $\eta_{\rm
w}$ requires an increase in $\theta_{\rm w}$ by a similar factor.

As was already noted in \citet{PKG05}, another realization of this type
of an outflow is potentially provided by a relativistic, baryon-poor
jet, which is driven electromagnetically along disk-anchored magnetic
field lines that thread the horizon of a rotating black hole, and which
is ``contaminated'' by neutron diffusion from a baryon-rich disk
wind. In the original version of this scenario, which was proposed by
Levinson \& Eichler (\citeyear{LE93}, \citeyear{LE03}; see also
\citealt{VPL03}) and recently studied numerically by \citet{M05a,M05b},
the narrow and wide components correspond to the baryon-poor and
baryon-rich outflows, respectively. However, the predicted Lorentz
factor of the disk wind is too low to be consistent with the rather high
value of $\eta_{\rm w}$ inferred from our model fits. An alternative
possibility is that the wide component corresponds to a
hydromagnetically accelerated baryon-rich disk outflow of the type
modeled by \citet{VK03a,VK03b}, whereas the narrow component corresponds
(as in the Levinson \& Eichler picture) to a baryon-poor
\citet{BZ77}-type outflow.\footnote{It is in principle also conceivable
that the baryon-poor outflow could develop an internal structure that
would correspond to the two outflow components considered here, but
there are at present no quantitative results to support this
conjecture.}  It should be noted that in either one of the above
two-component jet realizations the $\gamma$-ray emitting component is
associated with an initially Poynting-dominated outflow. This could in
principle make it possible to account for the relatively high emission
efficiency that the {\it Swift}\/ results seem to imply (see
\S~\ref{sec:conclusion}).

\section{Summary and Conclusions}
\label{sec:conclusion}

We have shown that the $\gamma$-ray efficiency implied by the {\it
Swift}\/ observations is model-dependent and can vary over a wide
range (from typical values of $\epsilon_\gamma\sim 0.9$ or higher to
$\epsilon_\gamma\sim 0.1$ or lower) depending on the adopted model
assumptions. The $\gamma$-ray efficiency has been expressed in terms
of observable quantities (see eqs. [\ref{obs_ratio}] and
[\ref{eps_g_fk1}]), namely $\kappa$ and $f$, where $\kappa$ relates
the $\gamma$-ray emission to the late-time afterglow emission (and was
therefore available in the pre-{\it Swift}\/ era) and $f$ relates the
early- and late-time afterglow energies (and therefore became
available only with the launching {\it Swift}). We have shown that
there is a large uncertainty in the values of both $\kappa$ and $f$,
which translates into a corresponding uncertainty in the value of the
$\gamma$-ray efficiency, $\epsilon_\gamma$.  In the following
discussion we make the conservative assumption that we observe most of
the emitted energy in $\gamma$-rays ($\epsilon_{\rm obs,GRB}\approx
1$); if a significant fraction of the total radiated energy is emitted
outside of the observed photon energy range (e.g., at higher energies)
then this would increase the required value of $\epsilon_\gamma$ (see
equation [\ref{eps_g_fk1}]).

The kinetic energy of the afterglow shock has been estimated in
the pre-{\it Swift}\/ era using broad-band afterglow fits for a
small number of GRB sources that had the best available broad-band
afterglow data, yielding a typical value of $\kappa\sim 1$ with a
large scatter of almost an order of magnitude (see
\S~\ref{sec:pre_swift}). Substituting the values of the
microphysical parameters ($\epsilon_e$, $\epsilon_B$ and $p$) that
were derived from these fits into our equations for $E_{\rm
k,iso}$ generally yielded somewhat lower values of $\kappa$.
Finally, using our equations with fiducial values of the
microphysical parameters ($\epsilon_e=0.1$, $\epsilon_B=0.01$, and
$p=2.2$) gives a typical value of $\kappa\sim 0.1-0.2$, both for
the pre-{\it Swift}\/ and the {\it Swift}\/ GRB samples that we
use, with a moderate scatter. Specifically, for our pre-{\it
Swift}\/ ({\it Swift}) sample, $\langle\log_{10}\kappa\rangle =
-0.75$ ($-0.82$) corresponding to $\kappa = 0.18$ ($0.15$) and
$\sigma_{\log_{10}\kappa}= 0.60$ ($0.63$). Obviously, the choice
of fiducial values for the microphysical parameters is somewhat
arbitrary and affects the resulting value of $\kappa$. Higher
values of the microphysical parameters $\epsilon_e$ and
$\epsilon_B$ (e.g., $\epsilon_e \approx 0.3$, $\epsilon_B \approx
0.08$) are required in order to obtain an average value of
$\log\kappa$, using our equations, similar to that derived from
pre-{\it Swift}\/ broad-band afterglow fits.  Altogether, there is
almost an order-of-magnitude uncertainty in the typical value of
$\kappa$ (which ranges from $\sim 0.1-0.2$ to $\sim 1$).

Even if we adopt the high typical values of $\kappa$ ($\sim 1$)
inferred from pre-{\it Swift}\/ afterglow fits, it is important to
keep in mind that these values have been estimated on the basis of
the standard assumptions of afterglow theory. Changing these
assumptions would modify the inferred value of $\kappa$ for the
same fits. For example, as pointed out by \citet{EW05}, if only a
fraction $\xi_e<1$ of the electrons are accelerated to
relativistic energies, then there is a degeneracy where the same
observable quantities are obtained for $\epsilon_e\to\xi_e\epsilon_e$,
$\epsilon_B\to\xi_e\epsilon_B$, $n_{\rm ext}\to\xi_e^{-1}n_{\rm
ext}$, and $E_{\rm k,iso}\to\xi_e^{-1}E_{\rm k,iso}$. Since this
increases the inferred value of $E_{\rm k,iso}$ by a factor of
$\xi_e^{-1}$, $\kappa$ is reduced by the same factor in comparison
with the estimate from the standard theory (which uses $\xi_e=1$).

An alternative way of reducing the inferred value of $\epsilon_\gamma$
was proposed by \citet{PKG05} in the context of the pre-{\it Swift}\/
observations. Specifically, they considered a two-componet outflow
model with parameters that effectively corresponded to the parameter
$f$ having a value $< 1$. This parameter choice appears to be
inconsistent with {\it Swift}'s detection of an early flattish decay
phase in the X-ray light curve, which, when interpreted in the context
of the standard afterglow theory as arising from a gradual increase
with time of $E_{\rm k,iso}$, typically implies $f\sim 10$ (and, in
some cases, values of $f$ that are as high as $\sim 10^2-10^3$). It is
worth noting in this connection that the existence of a two-component
GRB jet model can be plausibly expected on various theoretical grounds
and has been suggested independently on the basis of fits to several
pre-{\it Swift}\/ afterglows (see discussion in
\citealt{PKG05}). Furthermore, the {\it Swift}\/ observations by no
means rule out this model, although they can be used to constrain its
parameters. We have illustrated this fact through the fit to the X-ray
light curve of GRB 050315 that we performed in \S~\ref{sec:2comp}
within the framework of this model. This fit has yielded plausible
ranges for the kinetic energies and opening angles of the two
components as well for as the Lorentz factor of the dominant (wide)
component. A key conclusion from this fit is that the kinetic energy
of the wide component is much larger than that of the narrow one
($E_{\rm k,w}/E_{\rm k,n} \sim 10^2$). Physically, the narrow and wide
components could conceivably correspond to a baryon-poor,
black-hole--driven outflow \citep{LE93,LE03} and a baryon-rich,
disk-driven outflow \citep{VK03a,VK03b}, respectively, although this
remains to be demonstrated. We also note that the two-component jet
parameters derived in \S~\ref{sec:2comp} were based on standard
assumptions; they could change if the underlying assumptions
(involving, for example, the values and time constancy of the
microphysical parameters) were altered.

In conjunction with the $\kappa\sim 1$ estimates of the standard
broad-band afterglow fits, values of $f \gtrsim 10$ imply $\gamma$-ray
radiative efficiencies $\epsilon_\gamma\gtrsim 0.9$.  Such high
efficiencies would be essentially impossible to achieve in any scheme,
such as the internal-shocks model, that tapped the kinetic energy of
the outflow for radiation. An alternative possibility that has been
discussed in the literature is the direct transfer of Poynting flux
\citep[which evidently is also a major contributor to the flow
acceleration --- e.g.,][]{DS02,VK03a} into nonthermal radiation
\citep[e.g.,][]{U94,T94}. It is at present unclear how to assess the
efficiency of this process. There are two generic possibilities:
dissipative fronts and magnetic reconnection sites. The first option
corresponds to overtaking collisions of magnetically dominated
relativistic streams and is not expected to result in high radiative
efficiencies \citep[e.g.,][]{RL97,LvP97}. The second case would
require magnetic field orientation reversals and would most naturally
arise in a pulsar-type outflow from a rapidly rotating neutron star
\citep[e.g.,][]{SDD01}. In this case it is, however, still unclear
whether radiative efficiencies $\gtrsim 0.9$ could be attained even under
the most favorable assumptions about the field reconnection rate
\citep{DS02}, and it has in fact been suggested that the reconnection
rate might be self-limiting \citep{LK01}.

An ``intermediate'' situation could prevail if $f \sim 10$ but
$\kappa\sim 0.1$, reflecting the possibility that $\kappa$ was
overestimated by the pre-{\it Swift}\/ afterglow fits (perhaps because
some of the assumptions of the standard theory do not hold --- e.g.,
$\xi_e\sim 0.1$ rather than $\xi_e=1$). Alternatively, $\kappa$ could
be $\sim 1$ but $f=1$, corresponding to the early flattish decay phase
reflecting an increase with time of the X-ray afterglow efficiency
$\epsilon_{\rm X}$ (due, e.g., to $p$ being $<2$ or to an increase
with time of $\epsilon_e$ or $\epsilon_B$; see \S~\ref{sec:epsilon_X})
rather than an early increase in $E_{\rm k,iso}$. In either one of
these cases the inferred $\gamma$-ray radiative efficiency would be
reduced to $\epsilon_\gamma\sim 0.5$. Although this value is less
extreme than the estimate discussed in the preceding paragraph, it is
worth noting that it is still fairly restrictive for the
internal-shocks model, in which it could potentially be attained only
if {\it all}\/ of the following conditions (already summarized in
\citealt{PKG05}) are satisfied: (1) the ratio between the maximum and
minimum initial Lorentz factors of the ejected shells is large enough
($\gtrsim 10$); (2) the distribution of initial Lorentz factors is
sufficiently nonuniform; (3) the shells are approximately of equal
mass and their number is large enough ($\gtrsim 30$), and (4) the
fraction of the dissipated energy that is deposited in electrons and
then radiated away is sufficiently high ($\epsilon_{e,\rm
GRB}\epsilon_{\rm rad,GRB}\gtrsim 0.5$), with a similar constraint
applying to the fraction $\epsilon_{\rm obs,GRB}$ of the radiated
energy that is emitted as the observed $\gamma$-rays (see
\citealt{B00} and \citealt{KS01}).  Only if {\it both}\/ $f\sim 1$ and
$\kappa\sim 0.1$ were satisfied (which could occur, for example, if
$\epsilon_e$ or $\epsilon_B$ initially increased with time and $\xi_e$
were $\sim 0.1$) would the inferred value of $\epsilon_\gamma$ drop to
$\sim 0.1$ and be compatible with the values that are expected to
arise under less constrained circumstances in the internal-shocks model.

\acknowledgements

We are grateful to D. Eichler, E. Ramirez-Ruiz, and P. Kumar for many
useful discussions. This research was supported in part by the US
Department of Energy under contract number DE-AC03-76SF00515 (J.~G.),
by NASA Astrophysics Theory Program grant NAG5-12635 (A.~K.), as well
as by the US-Israel BSF and by the Schwartzmann University Chair
(T.~P.). J.~G. and A.~K. thank the Racah Institute of Physics at the
Hebrew University in Jerusalem for hospitality during the early phase
of this work.

\clearpage
\newcommand{\rb}[1]{\raisebox{1.5ex}[0pt]{#1}}
\begin{deluxetable}{llcccccccccccc}
\tabletypesize{\scriptsize} \rotate \tablecaption{Estimates of
$\kappa$ for pre-{\it Swift}\/ GRBs with known redshifts}
\tablewidth{624pt}
\tablehead{ \colhead{GRB} & & \colhead{$L_{\rm X,iso,10hr}^{\;\dagger}$} &
\colhead{$E_{\rm\gamma,iso}^{{\rm obs}\;\ddagger}$} & \colhead{$E_{\rm
k,iso,10hr}^{\;\P}$} & \colhead{$E_{\rm k,iso,10hr}^{\rm
(PK02)\;\spadesuit}$} & \colhead{$E_{\rm k,iso,10hr}^{\rm
(PK02)\;\clubsuit}$} & \colhead{$E_{\rm k,iso,10hr}^{\rm
(Y03)\;\heartsuit}$} & \colhead{$E_{\rm k,iso,1d}^{\rm
(Y03)\;\diamondsuit}$} & & \colhead{$\kappa^{\;\spadesuit}$} &
\colhead{$\kappa^{\;\clubsuit}$} &
\colhead{$\kappa_{\rm 10hr}^{\;\heartsuit}$} & \colhead{$\kappa_{\rm
1d}^{\;\diamondsuit}$} \\ \colhead{\#} &
\colhead{\rb{$z$}} & \colhead{$(10^{46}\,{\rm erg/s})$} &
\colhead{$(10^{52}\,{\rm erg})$} & \colhead{$(10^{52}\,{\rm erg})$} &
\colhead{$(10^{52}\,{\rm erg})$} & \colhead{$(10^{52}\,{\rm erg})$} &
\colhead{$(10^{52}\,{\rm erg})$} & \colhead{$(10^{52}\,{\rm erg})$} &
\colhead{\rb{$\kappa^{\;\P}$}} & \colhead{(PK02)} & \colhead{(PK02)}
& \colhead{(Y03)} & \colhead{(Y03)}}
\startdata
970228  & 0.695  & 0.682 & 1.416 & 22.2 & ---    & ---  & ---  & --- & 0.064 & -
--  & --- & --- & --- \\
970508  & 0.835  & 0.374 & 0.546 & 12.5 & 1.31   & 4.0 & 1.50 & 1.6 &
0.044 & $0.42^{\;\ast}$ & $0.14^{\;\ast}$ & 0.36 & 0.34 \\
970828  & 0.985  & 1.76  & 21.98 & 54.8 & ---    & ---  & ---  & --- & 0.40  & -
--  & ---  & --- & --- \\
971214  & 3.418  & 8.96  & 21.05 & 258  & ---    & ---  & ---  & --- & 0.082 & -
--  & ---  & --- & --- \\
980613  & 1.096  & 0.536 & 0.536 & 17.7 & ---    & ---  & ---  & --- & 0.030 & -
--  & ---  & --- & --- \\
980703  & 0.966  & 1.02  & 6.012 & 32.6 & ---    & ---  & 61.6 & 13  & 0.18  & -
--  & --- & 0.098 & 0.46 \\
990123  & 1.600  & 12.83 & 143.8 & 364  & 339    & 22   & ---  & --- &
0.40  & 0.42 & 6.4 & ---  & --- \\
990510  & 1.619  & 8.209 & 17.64 & 238  & $70^{\,\star}$ & 9.6 & ---  & --- &
0.074 & $\ 0.25^{\,\star}$ & 1.8 & --- & ---  \\
990705  & 0.840  & 0.123 & 25.60 & 4.35 & ---    & ---  & ---  & --- & 5.89  & -
--  & ---  & --- & --- \\
991216  & 1.02   & 18.32 & 53.54 & 510  & $\ \ $---$^{\,\star}$ & 9.9 & ---  & -
-- &
0.10 & $\ \ $---$^{\,\star}$ & 5.4 & ---  & --- \\
000210  & 0.846  & 0.183 & 16.93 & 6.35 & ---    & ---  & ---  & --- & 2.67  & -
--  & ---  & --- & --- \\
000926  & 2.037  & 7.169 & 27.97 & 209  & 36.7   & 3.2  & 272  & 15  &
0.13  & 0.76 & 8.7 & 0.10 & 1.86 \\
010222  & 1.477  & 13.79 & 85.78 & 389  & $19^{\,\star}$ & 16  & ---  & --- &
0.22  & $4.5^{\,\star}$ & 5.4 & --- & --- \\
011211  & 2.14   & 0.886 & 6.723 & 28.5 & ---    & ---  & ---  & --- & 0.24  & -
--  & ---  & --- & --- \\
020405  & 0.698  & 1.729 & 7.201 & 53.9 & ---    & ---  & ---  & --- & 0.13  & -
--  & ---  & --- & --- \\
020813  & 1.254  & 12.12 & 77.50 & 344  & ---    & ---  & ---  & --- & 0.23  & -
--  & ---  & --- & --- \\
021004  & 2.323  & 6.536 & 5.560 & 191  & ---    & ---  & ---  & --- & 0.029 & -
--  & ---  & --- & --- \\
\enddata
\tablecomments{
\label{pre-Swift}
    Estimates for $E_{\rm k,iso}(10\;{\rm hr})=E_{\rm k,iso,10hr}$ and
    $\kappa=E_{\rm\gamma,iso}^{\rm obs}/E_{\rm k,iso,10hr}$ for the
    GRBs with known redshift from the \citet{BFK03} and \citet{BKF03}
    samples. $^\dagger$ $L_{\rm X,iso,10hr} = L_{\rm X,iso}(10\;{\rm
    hr})$ from Table 2 of \citet{BKF03},
    $^\ddagger\;E_{\rm\gamma,iso}^{\rm obs}$ in the $20-2000\;$keV
    range form Table 2 of \citet{BFK03}, $^\P\;$calculated using
    eq. (\ref{E_k_iso}) with our fiducial values of the
    microphysical parameters: $\epsilon_e=0.1$, $\epsilon_B=0.01$, and
    $p=2.2$. $^\spadesuit\;$substituting the best fit values of the
    microphysical parameters from Table 2 of \citet{PK02} into
    eq. (\ref{E_k_iso}), $^\clubsuit\;$the value for $E_{\rm
    k,iso}(10\;{\rm hr})$ is taken to be $\approx 0.5E_{\rm k,iso,0}$,
    where $E_{\rm k,iso,0}$ is taken from \citet{PK02} and the factor
    of $\approx 0.5$ accounts for the average factor by which the
    energy is reduced relative to $E_{\rm k,iso,0}$ due to radiative
    losses (A. Panaitescu, personal communication), $^\ast\;$these
    values are for a fit to a uniform external density; PK02 get a
    significantly better fit to a wind density for which we derive
    $\kappa = 0.47$ and $0.17$ instead of $0.42$ and $0.14$,
    respectively, $^\star\;$in these cases \citet{PK02} find $p<2$
    which introduces an uncertainty through the extrapolation that is
    involved in the expression for the numerical coefficient in
    eq. (\ref{E_k_iso_bar}); for GRB 991216 PK02 find that the
    X-ray emission is from electrons with $\gamma_e>\gamma_{\rm max}$
    where $dN/d\gamma_e\propto\gamma_e^{-q}$ with $p<2<q$ so that we
    cannot readily substitute their results into our equations,
    $^\heartsuit\;$substituting the best fit values of the
    microphysical parameters from Table 1 of \citet{Y03} into eq.
    (\ref{E_k_iso}), $^\diamondsuit\;E_{\rm k,iso,1d}=E_{\rm
    k,iso}(1\;{\rm day})=E_{\rm\gamma,iso}^{\rm obs}/\kappa_{\rm 1d}$
    from Table 1 of \citet{Y03}.
    $^{\diamondsuit\diamondsuit}\;\kappa_{\rm
    1d}=E_{\rm\gamma,iso}^{\rm obs}/E_{\rm k,iso,1d}$. }

\end{deluxetable}

\clearpage
\begin{deluxetable}{llccccccc}
\tablecaption{Estimates of $f$ and $\kappa$ for {\it Swift}\/ GRBs
with known redshifts}
\tablehead{ & & \colhead{$L_{\rm X,iso,10hr}^{\;\dagger}$} &
\colhead{$E_{\rm\gamma,iso}^{{\rm obs}\;\ddagger}$} & \colhead{$E_{\rm
k,iso,10hr}^{\;\P}$} & & & & \\ \colhead{\rb{GRB \#}} &
\colhead{\rb{$z$}} & \colhead{$(10^{46}\,{\rm erg/s})$} &
\colhead{$(10^{52}\,{\rm erg})$} & \colhead{$(10^{52}\,{\rm erg})$} &
\colhead{\rb{$p^{\;\spadesuit}$}} &
\colhead{\rb{$\kappa^{\;\clubsuit}$}} &
\colhead{\rb{$f_{\rm
min}^{\;\heartsuit}$}} & \colhead{\rb{$f_{\rm max}^{\;\heartsuit}$}}}
\startdata
050126  & 1.29   & 0.12  & 2.2   & 6.86 & 3   & 0.055            & --- & --- \\
050315  & 1.949  & 16    & 18    & 699  & 2.1 & 0.034            & 11  & 29  \\
050318  & 1.44   & 0.60  & 3.9   & $\gtrsim 28.4$ & 2.1 & $\lesssim 0.18$  & 4.2
& 170 \\
050319  & 3.24   & 5.1   & 12.1  & $\gtrsim 118$  & 2.6 & $\lesssim 0.039$ & 12
& 76  \\
050401  & 2.90   & 9.8   & 137   & 433  & 2.1 & 0.41             & 5.6 & 14  \\
050408  & 1.236  & 1.4   & 2.9   & 38.5 & 2.3 & 0.058            & --- & --- \\
050416A & 0.6535 & 0.091 & 0.09  & 4.51 & 2.1 & 0.026            & 2.2 & 9.9 \\
050505  & 4.3    & 2.3   & 89    & $\gtrsim 59.0$ & 2.6 & $\lesssim 0.58$  & 19
& 1800 \\
050525A & 0.606  & 0.12  & 3.1   & $\gtrsim 3.96$ & 2.4 & $\lesssim 0.47$  & 2.1
& 5.9 \\
050603  & 2.821  & 1.1   & 126   & $\gtrsim 29.6$ & 2.4 & $\lesssim 2.57$  & ---
& --- \\
\enddata
\tablecomments{
\label{Swift}
The estimates for $f = E_{\rm k,iso,10hr}/E_{\rm k,iso,0}^{\rm obs}$
and $\kappa=E_{\rm\gamma,iso}^{\rm obs}/E_{\rm k,iso,10hr}$ for the
GRBs with known redshift from the \citet{Nousek05} sample;
$^\dagger\;L_{\rm X,iso,10hr}=L_{\rm X,iso}(10\;{\rm hr})$ in the
$2-10\;$keV range at (both the time and the photon energies measured
in the cosmological frame of the GRB) from Table 2 of
\citet{Nousek05}, $^\ddagger\;E_{\rm\gamma,iso}^{\rm obs}$ in the
$20-2000\;$keV range (in the GRB's cosmological frame) form Table 2 of
\citet{Nousek05}, $^\P\;$calculated using eq. (\ref{E_k_iso})
with $\epsilon_e=0.1$, $\epsilon_B=0.01$, and the values of $p$ from
this Table, $^\spadesuit\;$estimated using the measured spectral slope
in the X-rays \citep{Nousek05} and attempting to fit it into the range
$2<p<3$ if allowed within the errors on the spectral slope,
$^\clubsuit\;\kappa = E_{\rm\gamma,iso}^{\rm obs}/E_{\rm k,iso,10hr}$
estimated using the values from this Table, $^\heartsuit\;f_{\rm min}$
and $f_{\rm max}$ are taken from Table 3 of \citet{Nousek05}, and are
estimated using the measured X-ray flux at $t_{\rm break,1}$, and the
extrapolated X-ray flux at $T_{\rm GRB}$, respectively (see
\citealt{Nousek05} for details).}

\end{deluxetable}

\newpage
\begin{figure}
\plotone{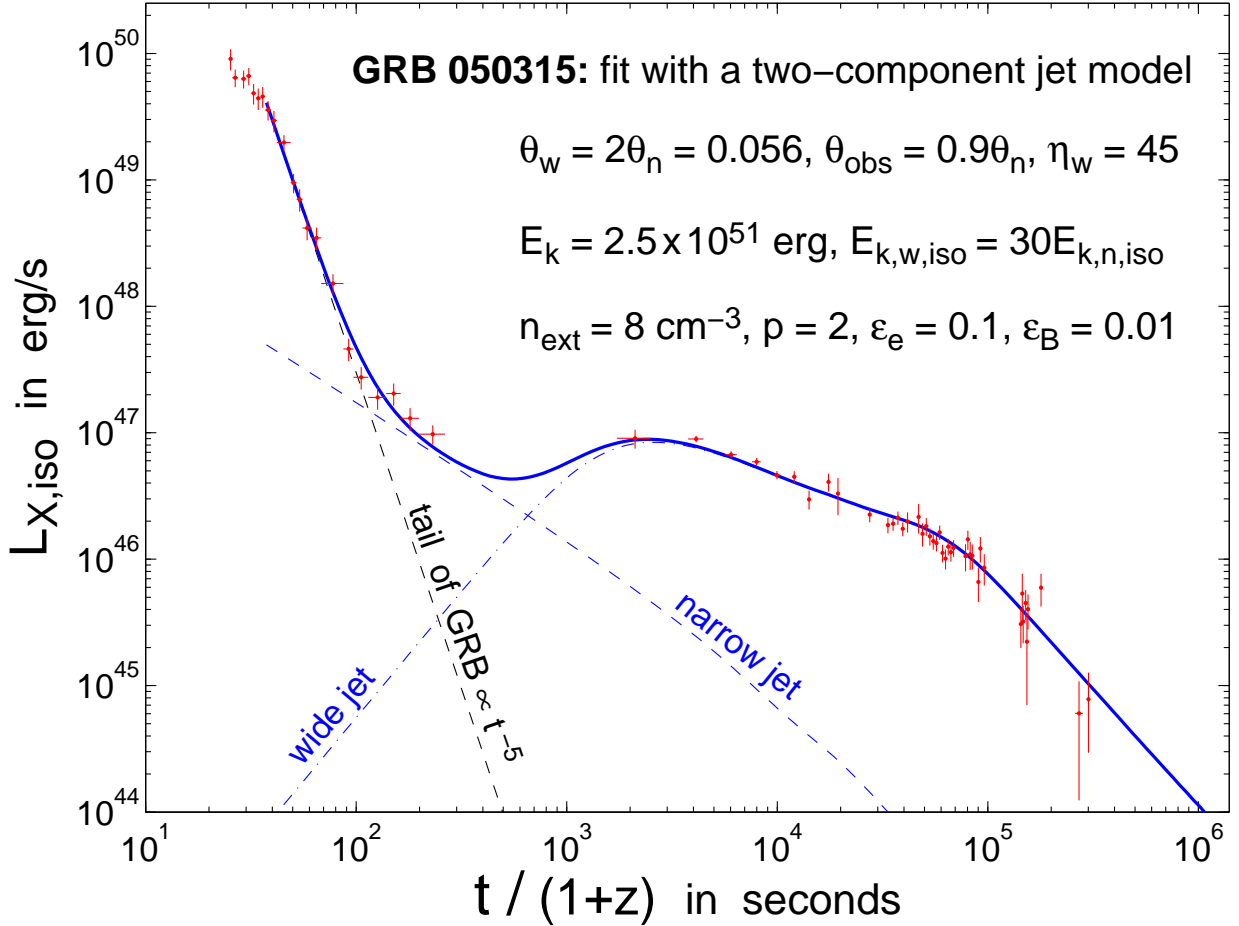}
\caption{Tentative fit to the X-ray light curve of GRB~050315
\citep[from][]{Nousek05} with the two-component jet model. The
numerical code used to calculate the light curve is essentially Model
1 of \citet{GK03}, which neglects the lateral spreading of the jet and
includes synchrotron self-Compton (SSC) emission. In addition to the
total light curve ({\it thick solid line}) we also show the separate
contributions of the different components: the tail of the prompt
emission ($\propto t^{-5}$), the narrow outflow, and the wide
outflow. Here $E_{\rm k} = E_{\rm k,w}+E_{\rm k,n}$ is the total
kinetic energy of the two components. The narrow and wide components
occupy the non-overlapping ranges $\theta<\theta_{\rm n}$ and
$\theta_{\rm n}<\theta<\theta_{\rm w}$, respectively, in the polar
angle $\theta$ (measured from the symmetry axis); $\theta_{\rm obs}$
is the viewing angle with respect to this axis.}
\label{2comp}
\end{figure}

\newpage
\begin{figure}
\plotone{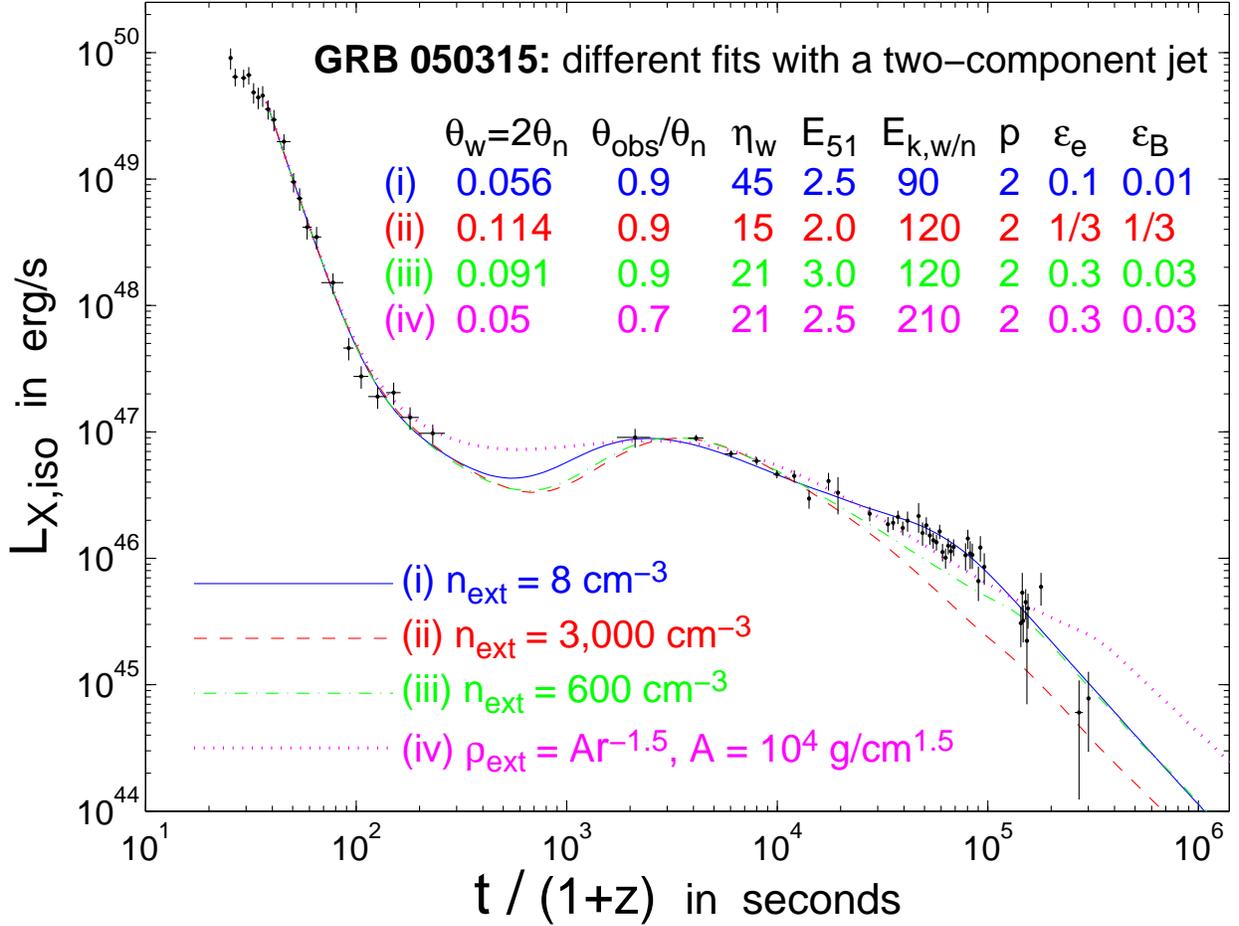}
\caption{The effects of varying the two-component jet model parameters
  with respect to those of the fit shown in Fig.~\ref{2comp}
  (reproduced in this figure by the {\it solid line}). Here
  $E_{51}=E_{\rm k}/(10^{51}\;{\rm erg})$ (with $E_{\rm k} = E_{\rm
  k,w}+E_{\rm k,n}$), whereas $E_{\rm k,w/n}$ denotes the ratio
  $E_{\rm k,w}/E_{\rm k,n}$ of the kinetic energies of the wide and
  narrow components.}
\label{2comp_diff}
\end{figure}

\end{document}